\documentclass[twocolumn]{autart}    

\usepackage{graphics}
\usepackage[dvips]{color}
\usepackage[pdftex]{graphicx}
\usepackage{multicol}
\usepackage{times}
\usepackage{amssymb}
\usepackage{amsmath}
\usepackage{harvard}
\citationstyle{dcu}
\usepackage{flafter}
\usepackage[center]{caption2}
\usepackage{indentfirst,latexsym,bm}
\usepackage{picinpar}
\usepackage[toc,page,title,titletoc,header]{appendix}       

\def\dmin{\mathop{\min}\limits}
\def\dmax{\mathop{\max}\limits}

\def\dsum{\displaystyle\sum}

\def\ba{\begin{array}}
\def\ea{\end{array}}
\def\ban{\begin{eqnarray*}}
\def\ean{\end{eqnarray*}}
\def\be{\begin{equation}}
\def\ee{\end{equation}}
\def\bna{\begin{eqnarray}}
\def\ena{\end{eqnarray}}

\def\nn{\nonumber}

\def\dref#1{(\ref{#1})}

\def\q{\quad}

\begin{document}

\begin{frontmatter}

\title{Distributed sampled-data control of nonholonomic multi-robot systems with proximity
networks\thanksref{footnoteinfo}} 

\thanks[footnoteinfo]{This work
was supported by the National Natural Science Foundation of China
under grants 61273221 and 61473189, the National Key Basic Research
Program of China (973 program) under grant 2014CB845302, and the
Australian Research Council's Discovery Projects funding scheme
under project DP130101658.}

\author[lzx]{Zhixin LIU} \ead{Lzx@amss.ac.cn},    \quad 
\author[wl]{Lin Wang} \ead{wanglin@sjtu.edu.cn},     \quad
\author[wjh]{Jinhuan Wang} \ead{jinhuan@hebut.edu.cn},     \quad
\author[ddy]{Daoyi Dong} \ead{daoyidong@gmail.com},     \quad
\author[hxm]{Xiaoming Hu}\ead{hu@kth.se}              

\address[lzx]{Key Laboratory of Systems and Control, Academy of Mathematics and Systems Science,
CAS, 100190, Beijing, China.}


\address[wl]{Department
of Automation, Shanghai Jiaotong University,
Shanghai 200240, P. R. China.}

\address[wjh]{ School of
Sciences, Hebei University of Technology, Tianjin 300401, P. R.
China.}

\address[ddy]{ School of Engineering and Information
Technology, University of New South Wales, Canberra, 2600 Australia.}

\address[hxm]{ Optimization and Systems Theory, KTH Royal
Institute of Technology, 10044 Stockholm, Sweden.}




\begin{abstract}                          
This paper considers the distributed sampled-data control problem of
a group of mobile robots connected via distance-induced proximity
networks. A dwell time is assumed in order to avoid chattering in
the neighbor relations that may be caused by abrupt changes of
positions when updating information from neighbors. Distributed
sampled-data control laws are designed based on nearest neighbor
rules, which in conjunction with continuous-time dynamics results in
hybrid closed-loop systems. For uniformly and independently
initial states, a sufficient condition is provided to
guarantee synchronization for the system without leaders. In order
to steer all robots to move with the desired orientation and speed,
we then introduce a number of leaders into the system, and
quantitatively establish the proportion of leaders needed to
track either constant or time-varying signals. All these conditions
depend only on the neighborhood radius, the maximum initial moving
speed and the dwell time, without assuming \emph{a prior} properties
of the neighbor graphs as are used in most of the existing
literature.

\end{abstract}

\end{frontmatter}

\section{Introduction}
Cooperative control of multi-robot/agent systems (MRS/MAS) has
generated wide interest for researchers in control and robotics
communities. Compared with a single robot, multiple robots can
cooperatively accomplish complicated tasks with the advantages of
high efficiency and robustness to the link failures. Over the last
decade,  MRS have wide applications in implementing a large number
of tasks ranging from coverage, deployment, rescue, to surveillance
and reconnaissance. Among these tasks, a basic one is to reach
synchronization, i.e., all robots reach the same state, which
actually has close connection with many important engineering
applications, such as rendezvous problem \cite{rend,Franices}, agreement problem \cite{agree}, distributed
optimization \cite{oz} and formation control \cite{formation}.

Recently, the synchronization problem of MAS has been extensively
studied in the literature where the neighbor relations are typically
modeled as graphs or networks.  For example,
 \citeasnoun{Jad} and \citeasnoun{ren}, respectively, studied the
first-order discrete-time MAS with undirected graphs and directed
graphs. \citeasnoun{murry} studied the MAS with
first-order continuous-time dynamics. The nonholonomic unicycle MRS
are investigated by  \citeasnoun{vision1} and \citeasnoun{vision2}. MAS with
nonlinear dynamics, time delays, and measurement noises are also
considered \cite{moreau,zhangya,wanglin,litao,shiguodong}.
In almost all existing results, the neighbor graphs are required to
satisfy certain connectivity assumptions for synchronization. How to verify or guarantee such conditions has been
a challenging issue. In order to maintain connectivity of dynamical
communication graphs, potential function methods are commonly used
when designing the distributed control laws
\cite{leader1,diamos,diamos1,connecprev}.

For a real world MRS, it is more practical that the dynamics of
the system are modeled in a continuous-time manner whereas the
control laws are designed based on the sampled-data information. The
sampled-data technique is of interest in many situations, such as
unreliable information channels, limited bandwidth, transport delay.
The synchronization of MAS with sampled-data control laws has been
studied \cite{sampling1,sampling2}, where the
neighbor graphs are also required to satisfy certain connectivity
assumptions. It is clear that the potential function techniques are
not applicable for the analysis of MAS with continuous-time dynamics
and sampled-data control, because connectivity of the networks might
be lost between sampling instants. How to analyze the
synchronization behavior of such kind of systems becomes more
challenging. In this paper, we first present a distributed
sampled-data algorithm for a group of nonholonomic unicycle robots
with continuous-time dynamics, and provide a comprehensive analysis
for the synchronization of the closed-loop hybrid system. In our
model, each robot has limited sensing and communication range, and
the neighbor relations are described by proximity networks. A dwell
time is assumed when updating information from neighbors,
implying that the control signals are kept constant between the
sampled instants and only updated at discrete-time instants.  With
such sampled-data information, our design of distributed control
laws based on nearest-neighbor rules will clearly result in a
hybrid closed-loop system, which is different from the case of
discrete-time MAS studied by \citeasnoun{tang2} and \citeasnoun{Auto}, and
is also different from the previous results given by
\citeasnoun{ifac} where the control law for the rotational speed is
designed using the continuous-time information.

For a multi-agent system, we may design a distributed algorithm to
guarantee synchronization of the system, but the synchronization
state is inherently determined by the initial states and model
parameters. In many practical applications, we expect that the
system achieves a desired synchronization state and we can treat
that state as a reference signal. The agents that have access to the
reference signal are referred to as leaders, and they can help steer
the MRS to the desired state. Although a large number of theoretical
analysis and results for the leader-follower model have been
provided, further theoretical investigation is still needed due to
some limitations in the existing theory: i) Similar to the
leaderless case, the neighbor graphs are required to be connected or
contain spanning trees to guarantee that the followers track the
reference signal  \cite{Jad,diamos1,toven,renwei2}, but there are few results to address how
to verify such conditions. ii) In order to guide all agents to
accomplish complicated tasks, such as tracking time-varying signals
and the containment control problem, a number of (not only one)
leaders need to be introduced into the system \cite{renwei2,nature,leaders1,leaders2}. However, quantitative theoretical results for the
number of leaders needed are still lack. Hence, this paper considers
also a multi-unicycle system with multiple leaders and presents some
new  quantitative results. The sampled-data information is used to
design the control laws for the followers and leaders. For the case
of the constant reference signal, we analyze the MRS with
heterogeneous agents where the leaders and followers have different
dynamics since the reference signal is only obtained by the leaders,
and quantitatively provide the proportion of leaders needed to track
the reference signal. In addition, we investigate the case where the
reference signal is dynamic but piecewise constant, and provide
quantitative results for the proportion of leaders needed to track a
slowly time-varying signal by analyzing the hybrid dynamics at each
stage.


The main contributions of this paper are summarized into the
following three aspects.
(i) For the leaderless case, we establish a sufficient condition,
imposed on the neighborhood radius, the dwell time and the maximum
moving speed, to guarantee synchronization of the nonholonomic
unicycles, which overcomes the difficulty of requiring \emph{a
prior} connectivity assumption on neighbor graphs used in most of the
existing results.
(ii) For the leader-follower model, we provide the proportion  of
leaders needed to guide all robots to track a reference signal which
can be constant or slowly time-varying. These quantitative results
illustrate that adding leaders is a feasible approach to guide MRS
to accomplish some complicated tasks.
(iii) For both the leaderless case and leader-following case, we
provide comprehensive analysis for nonlinear hybrid closed-loop
systems.  Different from the work of  \citeasnoun{tang2} and  \citeasnoun{Auto},
we need to estimate the synchronization rate of the continuous-time
variables (i.e., speed and orientation). Here the speed and
orientation are determined by the corresponding values at sampling
time instants and they are updated according to the states of
relevant neighbors, and the neighbors are defined via the positions
of all robots. Hence, the positions, orientations and moving speeds
of all robots are coupled. We deal with the coupled relationships by
combining the dynamical trajectories of the robots at discrete-time
instants with the analysis of continuous-time dynamics in sampling
intervals.

The rest of this paper is organized as follows. In Section
\ref{leaderless}, we present the problem formulation for a
leaderless model and provide the main result for synchronization. In
Section \ref{leader}, we first investigate the leader-following
problem where the leaders have constant reference signal, and
quantitatively provide the ratio of the number of leaders to the
number of followers needed to track the signal. We then extend our
result to the dynamical tracking where the leaders have time-varying
reference signal, and present some simulations to illustrate our
theoretical results. Concluding remarks are presented in Section
\ref{concluding}.


Notations:  For a vector $x\in \mathbb{R}^m$, $x^\prime$ denotes the
transpose of $x$, and $\|x\|$ denotes the 2-norm, i.e.,
$\|x\|=(x^{\prime}x)^{1/2}$. For a square matrix
$A=(a_{ij})_{n\times n}$, $\|A\|$ denotes the 2-norm of $A$, i.e.,
$\|A\|=\sqrt{\lambda_{\max}AA^\prime}$. For any two positive
sequences $\{a_{n}, n\geq 1\}$ and $\{b_{n}, n\geq 1\}$,
$a_{n}=O(b_{n})$ means that there exists a positive constant $C$
independent of $n$, such that $a_{n}\leq Cb_{n}$ for any $n\geq 1$;
$a_{n}=o(b_{n})$ (or ($a_n\ll b_n$)) means that
$\lim_{n\rightarrow\infty}\frac{a_{n}}{b_{n}}=0$; $a_n=\Theta(b_n)$,
if there exist two positive constants $C_1$ and $C_2$, such that
$C_1 b_n\leq a_n\leq C_2 b_n$.

\section{Leaderless Synchronization}\label{leaderless}


\subsection{Problem Formulation}

Consider a group of $n$ unicycle robots (or agents) moving in a
plane. For a robot $i$ $ (i=1, 2\cdots, n)$, the position of its
center at time $t$ $(t\geq 0)$ is denoted by $X_i(t)=(x_i(t),
y_i(t))^{\prime}\in \mathbb{R}^2$.  The orientation and moving speed
of each robot are affected by the states of its local neighbors. The
pair of two robots is said to be neighbors if their Euclidean
distance is less than a pre-defined radius $r_n$. We use
$\mathcal{N}_i(t)$ to denote the set of the robot $i$'s neighbors at
time $t$, i.e., \bna\label{10} \mathcal{N}_i(t)=\left\{j:
\Delta_{ij}(t)< r_n\right\},\ena where
$\Delta_{ij}(t)=\|X_i(t)-X_j(t)\|$ is the Euclidean distance between
robots $i$ and $j$. The cardinality of the set $\mathcal{N}_i(t)$,
i.e., the degree of the agent $i$, is denoted as $d_i(t)$. When the
robots move in the plane, the neighbor relations dynamically change
over time. We use graph $G_t=\{V, E_t\}$ to describe the
relationship between neighbors at time $t$, where the vertex set
$V=\{1, 2, \cdots, n\}$ is composed of all robots, and the edge set
is defined as $E_{t}=\{(i,j)\in V\times V: \Delta_{ij}(t)<r_n\}$.
The neighbor graphs are distance-induced, and also called geometric
graphs or proximity networks.


Let $\theta_i(t)$ and $v_i(t)$ denote the moving orientation and
translational speed of the $i$th robot  at time $t$. The dynamics of
the robots with nonholonomic constraint for pure rolling and
nonslipping is described by the following differential equations
(for $i\in V$), \bna \label{1.3} \left\{
              \begin{array}{ll}
                \dot{x}_i(t)=v_i(t)\cos\theta_i(t), & \hbox{} \\
                \dot{y}_i(t)=v_i(t)\sin\theta_i(t), & \hbox{} \\
                \dot{\theta}_i(t)=\omega_i(t), & \hbox{}\\
\dot{v}_i(t)=u_i(t), & \hbox{}
              \end{array}
            \right. \ena where $u_i(t)$ and $\omega_i(t)$ denote the acceleration and
rotational speed of the robot $i$ at time $t$. For robot $i$,
what we can control is its rotational speed $\omega_i(t)$ and its
acceleration $u_i(t)$, which is an extension from the standard unicycle
model where one controls the translational speed directly. We however
need to point out that with this simplified extension physical forces
affecting the angular motion,
such as side-slip forces and friction forces, are ignored.


For the feasibility of information processing by the robots, we
assume that the robots can only receive
information and design the control law at discrete-time instants
$t_0(=0), t_1, t_2, \cdots$. To simplify the analysis, we suppose
that the dwell time is the same and denoted by $\tau_n$, i.e.,
$t_{k+1}-t_k=\tau_n, k=0,1,\cdots$. At discrete-time instant $t_k$,
each agent is assumed to sense the relative speed and the relative
orientation of its neighbors. That is, for robot $i$, it receives
the following sampled-data information at time $t_k$,
$ \left\{v_j(t_k)-v_i(t_k),\
\theta_j(t_k)-\theta_i(t_k),\ j\in \mathcal{N}_i(t_k)\right\}.$

\begin{rem} From \dref{10}, we see that the neighbor set $\mathcal{N}_i(t)$ is determined by the positions
of all agents, so it is a continuous-time variable. The neighbor
relations might change abruptly when all robots are in motion. The
introduction of the dwell time avoids introducing chattering in the
abrupt changes caused by the evolution of positions.
\end{rem}

\begin{rem} For an agent, the relative
positions of its neighbors can be measured by e.g.,  the
geolocation and positioning technologies. Using the position
information and the orientation information of the agents, the
relative speed at the sampling instants can be calculated.  Moreover, the relative speed can also be estimated through observer-based methods by introducing reference robots, which is a different framework and falls into our future research. For the sake of
simplicity, we assume that each agent can receive the
relative speed of its neighbors in this paper.
\end{rem}

The objective of this section is to design the distributed control
law for the nonholonomic multi-robot system \dref{1.3} based on the
sampled-data information, such that the closed-loop
system becomes synchronized in both orientation and moving speed.
Here by synchronization we mean that there exists a common vector
$(v, \theta)$, such that for all $i\in V$, we have $
\lim_{t\rightarrow\infty}v_i(t)=v$ and $
\lim_{t\rightarrow\infty}\theta_i(t)=\theta.$

For robot $i$, we design the distributed control law according
to the widely used nearest-neighbor rule for $t\in [t_{k}, t_{k+1})$
($k=0,1,\cdots$),\bna \label{1.7} \left\{
                  \begin{array}{ll}
                     u_i(t)=\frac{1}{\tau_n d_i(t_k)}\sum_{j\in\mathcal{N}_i(t_k)}(v_j(t_k)-v_i(t_k)), & \hbox{} \\
                    \omega_i(t)=\frac{1}{\tau_n d_i(t_k)}\sum_{j\in\mathcal{N}_i(t_k)}(\theta_j(t_k)-\theta_i(t_k)), & \hbox{}
                  \end{array}
                \right.\ena where $d_i(t_k)$ is the degree of robot
$i$ at discrete-time $t_k$. Substituting \dref{1.7} into \dref{1.3},
we obtain the following hybrid closed-loop dynamical system:\bna &&
\label{2.3} \left\{
  \begin{array}{ll}
    \dot{x}_i(t)=v_i(t)\cos\theta_i(t), & \hbox{} \\
                \dot{y}_i(t)=v_i(t)\sin\theta_i(t), & \hbox{} \\
    \dot{\theta}_i(t)=\frac{1}{\tau_n d_i(t_{k})}\sum_{j\in\mathcal{N}_i(t_{k})}(\theta_j(t_k)-\theta_i(t_k)), &
\hbox{}\\
\dot{v}_i(t)= \frac{1}{\tau_n
d_i(t_k)}\sum_{j\in\mathcal{N}_i(t_k)}(v_j(t_k)-v_i(t_k)). &
\hbox{}\\
  \end{array}
\right.  \ena  Thus, we have for $t\in [t_k, t_{k+1}]$,
\bna\label{1.4}
&&\theta_i(t)=\left(1-\frac{t-t_k}{\tau_n}\right)\theta_i(t_k)+\frac{t-t_k}{\tau_n
d_i(t_k)}\sum_{j\in \mathcal{N}_i(t_{k})}\theta_j(t_k),\q \\
&&\label{2.4.1}v_i(t)=\left(1-\frac{t-t_k}{\tau_n}\right)v_i(t_k)+\frac{t-t_k}{\tau_n
d_i(t_k) }\sum_{j\in \mathcal{N}_i(t_{k})}v_j(t_k).\q \ena In
particular, at discrete-time instant $t=t_{k+1}$, the orientation
and moving speed evolve according to the following equations:  \bna
&&\label{4.5} \theta_i(t_{k+1})=\frac{1}{d_i(t_k)}\sum_{j\in
\mathcal{N}_i(t_{k})}\theta_j(t_k),\\&&\label{4.6}
v_i(t_{k+1})=\frac{1}{d_i(t_k)}\sum_{j\in
\mathcal{N}_i(t_{k})}v_j(t_k).\ena
In order to investigate the synchronization behavior of the hybrid
closed-loop system \dref{2.3}, we need to analyze the discrete-time
system at sampling time instants and the continuous-time system in
the sampling intervals simultaneously. For the discrete-time system
\dref{4.5} and \dref{4.6}, the algebraic properties of neighbor
graphs play a key role. Denote the Laplacian matrix of the graph
$G_t$ as $L(t)$. The normalized Laplacian matrix is defined as
$\mathcal{L}(t)= T^{-1/2}(t)L(t)T^{-1/2}(t)$, where the degree
matrix is defined as $T(t)=\hbox{diag}(d_1(t), d_2(t), \cdots,
d_{n}(t))$. The matrix $\mathcal{L}(t)$ is non-negative definite,
and 0 is one of the eigenvalues. Thus, we can arrange the
eigenvalues of $\mathcal{L}(t)$ according to such a non-decreasing
manner $0=\lambda_0(t)\leq \lambda_1(t)\leq \cdots\leq
\lambda_{n-1}(t)$. The spectral gap of graph $G_t$ is defined as $
\overline{\lambda}(t)=\max\{|1-\lambda_1(t)|,
|1-\lambda_{n-1}(t)|\}.$

Note that the dynamical behavior of all agents is determined by the
configuration formed by the initial states of the agents and
model parameters including neighborhood radius, the initial speed
and dwell time.  It is clear that there are numerous possibilities
for the initial configuration of the agents. If we do not impose any
assumption on the initial states of the agents, then we can only
carry out our analysis based on the worst case and the corresponding
results are considerably conservative. To solve this issue, we
introduce the following random framework, which accounts for a
natural setting on the initial states of all robots. In this
section, we consider the synchronization problem of the closed-loop
system \dref{2.3} under the following assumption, and aim to
establish synchronization conditions without relying on the
dynamical properties of neighbor graphs as are used in most
literature.

\begin{assum}\label{initial} 1) The positions, orientations and speeds
of all robots at the initial time instant are mutually independent;
2) For all robots, the initial positions are uniformly and
independently distributed (u.i.d.) in the (normalized) unit square
$[0,1]^2$; The initial headings are u.i.d. in $[-\pi, \pi)$; The
initial speeds are u.i.d. in the interval $[0, v_n]$.
\end{assum}

\begin{rem}\label{initialremarl}Under Assumption \ref{initial}, the initial
neighbor graph is called a random geometric graph (RGG), whose
properties are well investigated by \citeasnoun{geometric}. However, from
\dref{2.3}, the independency between positions of all agents will be
destroyed for $t=t_1, t_2, \cdots$. Thus, the properties concerning
the connectivity of static RGG can not be used.
\end{rem}

\begin{rem} Denote the sample spaces of the
position, orientation and speed as $\Omega_1=[0, 1]^2$,
$\Omega_2=[-\pi, \pi]$ and $\Omega_3=[0, v_n]$. By Assumption
\ref{initial}, our synchronization problem is discussed on the
sample space $\Omega=\underbrace{(\Omega_1\times\cdots\Omega_1)}_n
\times\underbrace{(\Omega_2\times\cdots\Omega_2)}_n
\times\underbrace{(\Omega_3\times\cdots\Omega_3)}_n$, where $\times$
denotes the Cartesian product.
\end{rem}
Divide the unit square $[0,1]^2$ into $M_n$ equally small squares
labeled as $1, 2,\cdots, M_n$, where $M_n=\lceil
\frac{1}{a_n}\rceil^2$ with $a_n$ satisfying $\sqrt{\log n/n}\ll
a_n\ll1$. Denote $N_k$ $(1\leq k\leq M_n)$  as the number of agents
in the corresponding small square. Introduce the sets \bna &&\label{square}B_n=\{\omega:\max_{1\leq j\leq
M_n}N_j=na_n^2(1+o(1))\}, \\ &&\label{theta0}\Theta_n=\big\{\omega: \max_{i\in V}\big|\sum_{j\in
\mathcal{N}_i(0)}\theta_j(0)\big|=O(f_n)\big\},\\&&\label{v}
\Gamma_n=\big\{\omega: \max_{i\in V}\big|\sum_{j\in
\mathcal{N}_i(0)}(v_j(0)-\frac{v_n}{2})\big|=O(v_n
f_n)\big\}, \ena
where $f_n=\sqrt{nr_n^2\log n}$. Using Lemma 7 given by
\citeasnoun{Auto}, the event $B_n$ happens with the probability
$\mathbb{P}(B_n)=1$ for large $n$, where $\mathbb{P}(\cdot)$ denotes
the probability of an event. By Assumption \ref{initial} and the
methods used in Lemma 9 given by \citeasnoun{Auto}, for large $n$, $
\mathbb{P}\left(\Theta_n \big| B_n\right)=1$ and $
\mathbb{P}\left(\Gamma_n \big| B_n\right)=1$ hold true. Furthermore, by the
independency of the orientations and moving speeds for the given
agents, it is clear that for large $n$,
$\mathbb{P}\left(\Theta_n\bigcap \Gamma_n \bigcap B_n\right)=1$. In
the following, we will investigate the dynamical behavior of all
agents on the set $\Theta_n\bigcap \Gamma_n \bigcap B_n$, and we
omit the words ``almost surely" (a.s.) for simplicity. Under the condition that the neighborhood radius
satisfies $\sqrt[6]{\log n/n}\ll r_n\ll 1$,  the maximum and minimum initial degrees satisfy the following equalities
\bna\label{degree}
 d_{\max}=n\pi r_n^2(1+o(1)); \
d_{min}(0)=\frac{n\pi r_n^2}{4}(1+o(1)).\ena In fact, for the case where the neighborhood radius independent of $n$, the degrees of the agents can also be estimated by similar methods.
The detailed analysis for the estimation of the initial degrees can be found in Theorem 2 given by
\citeasnoun{tang2} for details.

\subsection{Main Results}

We rewrite the orientation and speed update equations \dref{1.4} and
\dref{2.4.1} into the following matrix form for $ t\in [t_k,
t_{k+1}]$, \bna &&
\label{10.1}\theta(t)=\left(1-\frac{t-t_k}{\tau_n}\right)\theta(t_k)
+\frac{t-t_k}{\tau_n}P(t_k)\theta(t_k),\\&&
\label{10.2}v(t)=\left(1-\frac{t-t_k}{\tau_n}\right)v(t_k)
+\frac{t-t_k}{\tau_n}P(t_k)v(t_k). \ena Correspondingly, the
equations \dref{4.5} and \dref{4.6} are rewritten into the following
form, \bna && \label{10.3}\theta(t_{k+1})=P(t_k)\theta(t_k),\\&&
\label{10.4}v(t_{k+1})=P(t_k)v(t_k),\ena where the average matrix
$P(t_k)=(p_{ij}(t_k))$ is defined as: $p_{ij}(t_k)=\frac{1}{d_{i}(t_k)}$ if $(i,j)\in E_{t_k}$, and $p_{ij}(t_k)=0$ otherwise.

A known result for synchronization of the system \dref{10.3} and
\dref{10.4} can be stated as follows. If the neighbor graphs
$G(t_k)\ (k\geq 0)$ are connected, then all agents  move with
the same orientation and with the same speed eventually (cf.,
\citeasnoun{Jad}). The synchronization condition of the system \dref{10.3}
and \dref{10.4} is imposed on the dynamical properties of neighbor
graphs. However, for the system under consideration, the neighbor
graphs are defined via the positions of all agents. A comprehensive
analysis for the system \dref{2.3} should include how the change of
the positions of all robots affects the dynamical properties of
neighbor graphs, which brings challenges for our investigation.

Intuitively, for uniformly distributed agents, if the two agents are
not neighbors at the initial time instant, then they are still not
neighbors  with high probability as the system evolves. That is, if
the initial neighbor graph is disconnected, then it is hard to
obtain the connectivity of graph $G(t_k)$ for $k\geq 1$. In order to
reach synchronization, the connectivity of the initial neighbor
graphs is needed.  When the population size of the
agents increases, the number of neighbors of each agent will also
increase. Thus,  the interaction radius can be allowed to decay with
the number of agents to guarantee connectivity of the initial
neighbor graph. The properties of such graphs have been widely
investigated in the fields of wireless sensor networks and random
geometric graphs (cf., \citeasnoun{kumar}, \citeasnoun{geometric}). Moreover,
the change of neighbor graphs at the time instants $t_k (k>0)$ is
positively correlated with the moving speed and the dwell time.
Hence, the connectivity of the dynamical neighbor graphs at all
discrete-time instants may be preserved by assuming that the moving
speed and the dwell time are suitably small.

We now establish synchronization conditions for the hybrid system
\dref{2.3} depending only on the neighborhood radius, moving speed
and the dwell time. The states of the agents including
positions, orientations, and speeds of the system \dref{2.3} are
continuous-time variables. In order to investigate the
synchronization behavior of the system, we need to estimate  the
synchronization rate of the continuous-time variables $\theta_i(t)$
and $v_i(t)$. Moreover, by \dref{10.1} and \dref{10.2} $\theta_i(t)$
and $v_i(t)$ are affected by the discrete-time variables
$\theta_i(t_k)$ and $v_i(t_k)$ that are updated by the orientations
and speeds of relevant neighbors at $t_{k-1}$. While the neighbors
are defined via the positions of all robots. The positions,
orientations and moving speeds of all robots are coupled, which will
be handled by resorting to the mathematical induction and analyzing
the dynamics of the system. Thus, the comprehensive analysis
combines the dynamical trajectories of all agents at discrete
sampling time instants with the continuous time dynamics in sampling
intervals. The main theorem is stated as follows.

\begin{thm}\label{propo}  If the
neighborhood radius $r_n$, the maximum initial speed $v_n$, and the
dwell time $\tau_n$ satisfy the conditions $
\sqrt[6]{\frac{\log n}{n}}\ll r_n \ll 1$ and $v_n\tau_n\leq
c^{\prime}\frac{\eta_n r_n^3}{\log n}$, where $c^{\prime}$ is a
positive constant satisfying $0< c^{\prime}\leq \frac{1}{144}$ and
$\eta_n$  is taken as $\eta_n=c r_n^2$ with $c$ satisfying $0<c\leq
\frac{1 }{144\cdot 320}$. Under Assumption \ref{initial}, the system
\dref{2.3} reaches synchronization in orientation and speed almost
surely for large $n$.
\end{thm}

\textbf{Proof:} It is clear that under the condition for
the neighborhood radius, the initial neighbor graph is connected
with probability one (cf., \citeasnoun{kumar}). If the connectivity of the
neighbor graphs can be preserved, then synchronization can be
reached. Hence, a key step is to show that the distance between any
two robots $i$ and $j$ at any sampling instant $t_k$ ($k\geq 0$)
satisfies the following inequality, \bna\label{1.5}
&&|\Delta_{ij}(t_k)-\Delta_{ij}(0)|\leq \eta_n r_n. \ena We use
mathematical induction to prove \dref{1.5}. It is obvious that the
inequality \dref{1.5} holds for $k=0$. We assume that \dref{1.5}
holds for all $l\leq k_0$, and prove that it is true for $k_0+1$.

By the position update equation in \dref{2.3}, it is clear that the the distance between
any two robots $i$ and $j$ at contiguous discrete-time instants
satisfies \bna\label{distancerecur}
&&|\Delta_{ij}(t_{k+1})-\Delta_{ij}(t_k)|\leq
2\int_{t_k}^{t_{k+1}}\delta v(t)dt\nn\\&&\hskip 1.6cm +2\max_{1\leq i\leq
n}|v_i(t_k)|\int_{t_k}^{t_{k+1}}\delta \theta(t)dt,\   k\geq 1,\ena
where $\delta \theta(t)=\max_{ i, j\in
V}|\theta_i(t)-\theta_{j}(t)|$ and $\delta v(t)=\max_{i, j\in
V}|v_i(t)-v_{j}(t)|$ denote the dissimilarity of the orientation and
the speed between the two agents. In particular, the distance at time $t_1$
satisfies \bna\label{initialdistance}
|\Delta_{ij}(t_1)-\Delta_{ij}(0)|&\leq& 2\Big(\int_{0}^{t_1}\delta
v(t) dt+2\max_{1\leq i\leq n}v_i(0)\tau_n\Big)\nn\\&\leq& 8v_n\tau_n.\ena
The proof of \dref{distancerecur} is put in
\ref{prooflemma}, and the inequality \dref{initialdistance} can be
obtained by following the proof of \dref{distancerecur}.

By \dref{distancerecur}, we see that the analysis of the distance
depends on the convergence properties of $\delta v(t)$ and $\delta
\theta(t)$. We first estimate the convergence rate of $\delta
v(t)$. By the induction assumption, we have for $l\leq
k_0$\bna\label{average} \|P(t_l)-P(0)\|\leq80\eta_n(1+o(1)),\ena
whose proof is presented in  \ref{prooflemma}.  Using
 \dref{average}  and Lemma 2 given by \citeasnoun{tang2}, we have
for $1\leq l\leq k_0$ \bna\label{deltav} &&\delta v(t_{l+1})\leq\nn\\&&
\sqrt{2}\kappa\left(\overline{\lambda}(0)+\kappa\dmax_{1\leq s\leq
l}\|P(s)-P(0)\|\right)^{l}\|v(t_1)\|\leq\nn\\&&2\sqrt{2}\left(1-\frac{r_n^2}{144}(1+o(1))+160\eta_n\right)^l\|v(t_1)\|
(1+o(1))\nn\\&&\leq 2\sqrt{2}\left(1-\frac{r_n^2}{288}(1+o(1))\right)^l\|v(t_1)\|(1+o(1))\nn\\&&\triangleq
2\sqrt{2}\left(\widehat{\lambda}_n\right)^l\|v(t_1)\|(1+o(1)), \ena
where $\kappa=\sqrt{\frac{d_{\max}(0)}{d_{\min}(0)}}=2(1+o(1))$ by
using \dref{degree}, and $\overline{\lambda}(0)$ is the spectral gap
of the initial neighbor graph $G_0$. By Lemma 16 given by \citeasnoun{Auto},
$\overline{\lambda}(0)$ can be estimated as
$\overline{\lambda}(0)\leq 1-\frac{r_n^2}{144}$ . Hence, by the
speed update equation \dref{10.2}, we have for $1\leq l\leq k_0$
\bna\label{deltav2}&& \int_{t_l}^{t_{l+1}}\delta v(t)dt\leq
\int_{t_l}^{t_{l+1}}\Big[\big(1-\frac{t-t_l}{\tau_n}\big)\delta
v(t_l)\nn\\&&\hskip2.5cm+\frac{t-t_l}{\tau_n}\delta (v(t_{l+1}))\Big]dt \leq
2\sqrt{2}\left(\widehat{\lambda}_n\right)^{l-1}\nn\\&&\|v(t_1)\|
\int_{t_l}^{t_{l+1}}\Big(1-\frac{(t-t_l)(1-\widehat{\lambda}_n)}{\tau_n}\Big)dt(1+o(1))\nn\\
&&=
\sqrt{2}\left(1+\widehat{\lambda}_n\right)\left(\widehat{\lambda}_n\right)^{l-1}\|v(t_1)\|\tau_n(1+o(1)).\ena
On the other hand, using \dref{v} and \dref{degree}, we have
$\max_{i\in V}|v_i(t_1)|$ $ = \frac{v_n}{2}(1+o(1))$.
Thus, we obtain that for $l\geq 1$ \bna&& \int_{t_l}^{t_{l+1}}\delta
v(t)dt\leq 2\tau_n\max_{i\in V} v_i(t_l)\leq\tau_n v_n(1+o(1)).\ena
Set $ l_0=\min\{l:
\sqrt{2}(1+\widehat{\lambda}_n)(\widehat{\lambda}_n)^{l-1}\|v(t_1)\|\leq
v_n \},$ then we have $
l_0=\lceil\frac{\log\frac{v_n}{\sqrt{2}(1+\widehat{\lambda}_n)\|v_{t_1}\|}
}{\log \widehat{\lambda}_n}+1\rceil$ $\leq
\frac{\log\frac{v_n}{\sqrt{2}(1+\widehat{\lambda}_n)\|v_{t_1}\|}
}{\log \widehat{\lambda}_n}+2 .$ Thus, \bna\label{2.0}
&&\sum_{l=1}^{k_0}\int_{t_l}^{t_{l+1}}\delta v(t)dt\leq
\sum_{l=1}^{l_0-1}\tau_n\delta v(t_l)dt\nn\\&&+
\sum_{l=l_0}^{k_0}\sqrt{2}\left(1+\widehat{\lambda}_n\right)\left(\widehat{\lambda}_n\right)^{l-1}\|v(t_1)\|\tau_n(1+o(1))
\nn\\&\leq&
v_n\tau_n\left(l_0+\sum_{l=l_0}^{k_0}\left(\widehat{\lambda}_n\right)^{l-l_0}\right)(1+o(1))\nn\\&\leq&
v_n\tau_n\left(\frac{\log\frac{v_n}{\sqrt{2}(1+\widehat{\lambda}_n)\|v_{t_1}\|}
}{\log
\widehat{\lambda}_n}+2+\frac{1}{1-\widehat{\lambda}_n}\right)(1+o(1))\nn\\&\leq&\frac{v_n\tau_n(1+o(1))}{1-\widehat{\lambda}_n}\left(\log
\frac{\sqrt{2}(1+\widehat{\lambda}_n)\|v_{t_1}\|}{v_n}+3
\right)\\&=&\label{1.2}\frac{144 v_n\tau_n \log
n(1+o(1))}{r_n^2},\ena where the inequality $-\log x\geq 1-x$ for
$x\in (0,1)$ is used in \dref{2.0}.

By a similar analysis as that of \dref{deltav2}, we have for $1\leq
l\leq k_0$ \bna\label{3.1} &&
\int_{t_l}^{t_{l+1}}\delta\theta(t)dt\leq
\int_{t_l}^{t_{l+1}}\Big[\big(1-\frac{t-t_l}{\tau_n}\big)\delta
\theta(t_l)\nn\\&&\hskip 2cm+\frac{t-t_l}{\tau_n}\delta\theta(t_{l+1})\Big]
dt\leq
2\sqrt{2}(\widehat{\lambda}_n)^{l-1}\nn\\&&\|\theta(t_1)\|\int_{t_l}^{t_{l+1}}
\big(1-\frac{(t-t_l)(1+\widehat{\lambda}_n)}{\tau_n}\big)dt(1+o(1))\nn\\& &=
\sqrt{2}\left(1+\widehat{\lambda}_n\right)\left(\widehat{\lambda}_n\right)^{l-1}\|\theta(t_1)\|\tau_n(1+o(1)).\ena
Meanwhile,  we have for $l\geq 1$ \bna\label{3.1.1} &&
\int_{t_l}^{t_{l+1}}\delta\theta(t) dt\leq2\tau_n\max_{i\in
V}|\theta_i(t_1)|(1+o(1)).\ena By \dref{theta0} and \dref{degree},
we have $\max_{ i\in V}|\theta_i(t_1)|=O\left(\sqrt{\frac{\log
n}{nr_n^2}}\right)$. Similar to the analysis of \dref{1.2}, it is
clear that \bna\label{3.0}&&\sum_{l=1}^{k_0} \max_{1\leq i\leq
n}v_i(t_l)\int_{t_l}^{t_{l+1}}\delta\theta(t)dt\nn\\&\leq&\frac{v_n\tau_n\log
n}{2(1-\widehat{\lambda}_n)}\max_{1\leq i\leq
n}|\theta_i(t_1)|(1+o(1))\nn\\&=&\frac{144 v_n\tau_n\log n}{ r_n^2}\cdot
O\left(\sqrt{\frac{\log n}{nr^2_n}}\right).\ena
Substituting \dref{2.0} and \dref{3.0} into \dref{distancerecur}, we
can derive that the distance between agents $i$ and $j$ at time
$t_{k_0+1}$ satisfies \bna\label{5.2}
&&|\Delta_{ij}(t_{k_0+1})-\Delta_{ij}(0)|\leq\sum_{l=0}^{k_0}|\Delta_{ij}(t_{l+1})-\Delta_{l}(0)|\nn\\&\leq&
|\Delta_{ij}(t_1)-\Delta_{ij}(0)|+
2\sum_{l=1}^{k_0}\int_{t_l}^{t_{l+1}}\delta v(t)dt
\nn\\&&+2\sum_{l=1}^{k_0}\max_{1\leq i\leq
n}|v_i(t_l)|\int_{t_l}^{t_{l+1}}\delta \theta(t)dt.\nn\\&\leq&
\frac{288 v_n\tau_n \log n(1+o(1))}{r_n^2}\leq\eta_nr_n,\ena
where the conditions on the speed and the dwell time
are used in the last inequality. Using the mathematical induction,
we see that the inequality \dref{1.5} holds for all $k$. As a
consequence, we have for $k\geq 0$,\ban
\|P(t_k)-P(0)\|=80\eta_n(1+o(1)).\ean Thus, the inequality
\dref{deltav} holds for all $l$, i.e., \bna \label{1.1.3}&& \delta
v(t)\leq \delta v(t_k)\nn\\&\leq& \sqrt{2n}v_n\left(1-\frac{\pi
r_n^2(1+o(1))}{288}\right)^{k-1}(1+o(1))\nn\\&\rightarrow&0, \hskip
1.5cm \hbox{as}\ k\rightarrow\infty.\ena

By \dref{4.6}, it is clear that $\max_{i\in V}v_i(t_k)$ (resp.
$\min_{i\in V}$ $v_i(t_k)$) are non-increasing (resp.
non-decreasing) sequences. Thus, both the sequence $\max_{i\in
V}v_i(t_k)$ and $\min_{i\in V}v_i(t_k)$  have bounded limits as
$k\rightarrow\infty$. Moreover,  by \dref{1.1.3}, $\max_{i\in
V}v_i(t_k)$ and $\min_{i\in V}v_i(t_k)$ have the same limit, and the
translational speed $v_i(t)$ tends to the same value for all $i\
(1\leq i\leq n)$. By a similar analysis, we can prove that the
orientations of all agents tend to the same value as
$t\rightarrow\infty$. $\hfill\square$

\begin{rem} Generally speaking, the neighborhood radius has some physical
meanings in practical systems, e.g., reflecting the sensing ability
of sensors. The smaller the neighborhood radius is, the less the
energy consumption will be. From practical point of view, the
neighborhood radius should be as small as possible. However, for the
system under consideration, the larger the neighborhood radius, the
easier the system reaches synchronization. The neighborhood radius
for synchronization in Theorem \ref{propo} describes a tradeoff
between these two factors.
\end{rem}

\begin{rem}
Theorem \ref{propo} establishes scaling rates for the neighborhood
radius and moving speed for synchronization under Assumption
\ref{initial}. The conditions on these parameters can be adjusted
according to the practical demands. Assume that the agents are u. i.
d. in the square $[ 0, L_n]^2$. Let $ x_i^*(t)=\frac{1}{L_n} x_i(t)$ and $y_i^*(t)=\frac{1}{L_n} y_i(t) $. It is clear that $(x_i^*(0), y_i^*(0))$ is u. i. d. in the unit square $[0,1]^2$.  By \dref{2.3}, $x_i^*(t)$ and $y_i^*(t))^{\prime}$ are updated
 according to the equations $ \dot{x}_i^*(t)=\frac{1}{L_n}v_i(t)\cos\theta_i(t)$, and $\dot{y}_i^*(t)=\frac{1}{L_n} v_i(t)\sin\theta_i(t)$, respectively.
The neighbor relations can also be rewritten as $\mathcal{N}_i(t)=\left\{j: \Delta^*_{ij}(t)<
\frac{r_n}{L_n}\right\}$, where
$\Delta^*_{ij}(t)=\|X_i^*(t)-X_j^*(t)\|$. Following the proof line
of this paper, similar results for synchronization of the MRS can be
obtained just by replacing $v_n$ by $\frac{v_n}{L_n}$ and replacing
$r_n$ by $\frac{r_n}{L_n}$. By this and Theorem \ref{propo}, we can easily obtain the following result,
\begin{cor} Let the neighborhood radius $r$ and the initial maximum moving speed $v$ be two positive constants. If the dwell time satisfies $\tau_n\leq \widetilde{c}/\log n$ with $\widetilde{c}$ being a positive constant depending on $v$ and $r$, then under Assumption \ref{initial} the MAS reaches synchronization almost surely for large $n$.
\end{cor}
\end{rem}

\section{Leader-following of unicycle robots}\label{leader}

In Section \ref{leaderless}, we designed the distributed control law
for each robot using the sampled-data information, and provided
parameter conditions to guarantee synchronization of all robots. It
is clear that the resulting orientation and speed of unicycles are
determined by the initial states of all robots and model parameters.
For many practical applications, e.g., avoiding
collision with obstacles, following a given path, we may expect to
guide all robots towards a desired orientation and speed. To achieve
this goal, a cost efficient way is to introduce some special robots that have the reference
signal about the desired behavior of the whole system. These special
robots with the reference signal are called leaders, and other
ordinary robots are called followers. In this section, we study the
system composed of heterogeneous agents. The sampled-data control
laws of both leaders and followers are designed, and some
quantitative results for the proportion of leaders needed to track
the constant and time-varying signals are established.

\subsection{Problem Formulation}\label{problem}

We consider the system composed of  $n$ followers and $\rho_n=\lceil
n\alpha_n\rceil$ leaders, where $\alpha_n\ (\alpha_n\in(0,1])$ is
the ratio of the number of leaders to the number of followers. We
denote the follower set and leader set as $V_1=\{1, 2, \cdots, n\}$
and $V_2=\{n+1, n+2, \cdots, n+\rho_n\}$, respectively, and
$V=V_1\bigcup V_2$.

The dynamics of both leaders and followers is described by
\dref{1.3}. For the followers, they receive the sampled-data
information $\left\{v_j(t_k)-v_i(t_k),\
\theta_j(t_k)-\theta_i(t_k),\ j\in \mathcal{N}_i(t_k)\right\}$ at discrete-time instant $t_k$, and the
closed-loop dynamics is described by \dref{2.3}, i.e., for $i\in
V_1$ and $t\in [t_k, t_{k+1})$, \bna && \label{3.3} \left\{
  \begin{array}{ll}
    \dot{x}_i(t)=v_i(t)\cos\theta_i(t), & \hbox{} \\
                \dot{y}_i(t)=v_i(t)\sin\theta_i(t), & \hbox{} \\
    \dot{\theta}_i(t)=\frac{1}{\tau_n d_i(t_{k})}\sum_{j\in\mathcal{N}_i(t_{k})}(\theta_j(t_k)-\theta_i(t_k)), &
\hbox{}\\
\dot{v}_i(t)= \frac{1}{\tau_n
d_i(t_k)}\sum_{j\in\mathcal{N}_i(t_k)}(v_j(t_k)-v_i(t_k)). &
\hbox{}\\
  \end{array}
\right.  \ena Different from followers, the leaders can receive the
relative information of the desired orientation and the desired
speed, in addition to the relative translational speed and relative
orientation of their neighbors at discrete-time $t_k$
$(k=0,1,2,\cdots )$ with $\tau_n=t_{k+1}-t_k$. Thus, for a leader
robot $i\ (i\in V_2)$, it has the following information at time
instant $t_k$, $\{v_j(t_k)-v_i(t_k),\
\theta_j(t_k)-\theta_i(t_k), \ \overline{\theta}_0-\theta_i(t_k), \
v_n-v_i(t_k), \ j\in \mathcal{N}_i(t_k)\}$, where
$\overline{\theta}_0$ and $v_n$ are the desired orientation and the
desired speed, respectively. We adopt the control law for the
rotational speed $\omega_i(t)$ and the acceleration $u_i(t)$ of
leaders with the following form for $t\in [t_{k}, t_{k+1})$
\bna\label{3.6} && \omega_i(t)=\frac{1}{\tau_n}\Big\{\vartheta(\overline{\theta}_0-\theta_i(t_k))\nn\\&&\hskip 2cm+
\frac{1-\vartheta}{
d_i(t_k)}\dsum_{j\in\mathcal{N}_i(t_k)}(\theta_j(t_k)-\theta_i(t_k))\Big\},\\ \label{3.6.1}&& u_i(t)=\frac{1}{\tau_n}\Big\{\vartheta(v_n-v_i(t_k))\nn\\&&\hskip 2cm
+\frac{1-\vartheta}{
d_i(t_k)}\dsum_{j\in\mathcal{N}_i(t_k)}(v_j(t_k)-v_i(t_k))\Big\}, \ena
where the positive constant $0< \vartheta\leq1$ reflects the balance
between the expected behavior and local interactions with neighbors.
Substituting \dref{3.6} and \dref{3.6.1} into \dref{1.3}, we can obtain the
closed-loop dynamics for the headings and speeds of leaders for
$t\in [t_{k}, t_{k+1})$, \bna\label{4.4}
&&\theta_i(t)=(1-\frac{t-t_k}{\tau_n})\theta_i(t_k)\nn\\&&\hskip 1.8cm+\frac{t-t_k}{\tau_n
}\Big\{\vartheta\theta_0+\frac{1-\vartheta}{d_i(t_k)}\dsum_{j\in\mathcal{N}_i(t_k)}\theta_j(t_k)\Big\},\\
&&\label{2.4}v_i(t)=(1-\frac{t-t_k}{\tau_n})v_i(t_k)+\nn\\&&\hskip 1.8cm\frac{t-t_k}{\tau_n
}\Big\{\vartheta
v_n+\frac{1-\vartheta}{d_i(t_k)}\dsum_{j\in\mathcal{N}_i(t_k)}v_j(t_k)\Big\}.\ena
Particularly, for $t=t_{k+1}$, \bna &&\label{4.2}
\theta_i(t_{k+1})=\vartheta\theta_0+\frac{1-\vartheta}{d_i(t_k)}\dsum_{j\in\mathcal{N}_i(t_k)}\theta_j(t_k),\\&&
\label{3.9} v_i(t_{k+1})= \vartheta
v_n+\frac{1-\vartheta}{d_i(t_k)}\dsum_{j\in\mathcal{N}_i(t_k)}v_j(t_k).\ena
Note that after the leaders are added, the neighbor set of both
leaders and followers is composed of two parts: leader neighbors and
follower neighbors. For a robot $i\ (i\in V)$, we use
$\mathcal{N}_{i1}(t_k)$ and $\mathcal{N}_{i2}(t_k)$ to denote its
follower neighbor set and leader neighbor set at discrete-time
instant $t_k$, respectively. That is, $\mathcal{N}_{i1}(t_k)=\{j\in V_1:\ \|X_j(t_k)-X_i(t_k)\|< r_n\}$, and $\mathcal{N}_{i2}(t_k)=\{j\in V_2:\
\|X_j(t_k)-X_i(t_k)\|< r_n\}$. 
Denote the cardinality of the sets
$\mathcal{N}_{i1}(t_k)$ and $\mathcal{N}_{i2}(t_k)$ as $d_{i1}(t_k)$
and $d_{i2}(t_k)$, respectively. Thus, we have
$\mathcal{N}_i(t_k)=\mathcal{N}_{i1}(t_k)\bigcup\mathcal{N}_{i2}(t_k)$,
and $d_i(t_{k})=d_{i1}(t_{k})+d_{i2}(t_{k})$.

For the leader-follower model, if the union of neighbor graphs in
bounded time intervals contains a spanning tree rooted at the
leaders, then all robots will move with the desired orientation and
with the desired speed eventually. How to
guarantee the existence of the spanning tree is unresolved.

We aim at establishing the quantitative relationship between the
proportion of leaders and the neighborhood radius, moving speed and
dwell time such that all agents move with the expected orientation
$\overline{\theta}_0$ and the expected speed $v_n$ eventually.
It is clear that the initial distribution of leaders is
crucial for the proportion of leaders needed. For example, assume
that the leaders and the followers at the initial instant are
distributed in two disjoint areas, and the distance between these
two areas are large. At the initial time, the followers are not
affected by the reference signals of the leaders since the followers do
not have leader neighbors, and they evolve in a self-organized
manner. By \dref{4.4} and \dref{2.4}, the leaders converge to the desired
orientation and speed in a certain rate. When the system evolves,
the distance between the subgroups of leaders and followers becomes
larger and larger. As a result, the followers can not be guided to
the desired behavior no matter what the proportion of leaders is. In
this part, we proceed with our analysis under Assumption
\ref{initial} in which the uniform distribution of leaders makes it
possible to investigate the proportion of leaders for
synchronization.

\subsection{Main Results}

For $i\in V$, we denote $
\widetilde{\theta}_i(t)=\theta_i(t)-\overline{\theta}_0$ and
$\widetilde{v}_i(t)=v_i(t)-v_n.$ What we concern is the ratio of
leaders $\alpha_n$ needed such that  for all $i\in V$, we have
$\widetilde{\theta}_i(t)\rightarrow0$ and
$\widetilde{v}_i(t)\rightarrow0$ as $t\rightarrow\infty$. Different from the leaderless case where all agents have
the same closed-loop dynamics, the leaders and followers in the leader-following model have different closed-loop dynamics. Thus, we need to
analyze the synchronization of the system with heterogeneous agents.
Moreover, the orientation of each follower is affected by the
orientations of its neighbors including leader neighbors and
follower neighbors, and the orientation of each leader is affected
by the orientations of its leader neighbors and follower neighbors.
The coupled relationship makes the analysis of the leader-following model more challenging. We
first present a preliminary result for the convergence of orientations of leaders and followers.owers.


\begin{lem}\label{delta}If there exist positive constants $A$ and
$ \mu$, such that $\max_{i\in
V_1}\widetilde{\theta}_i(t_1)\leq A$, $\max_{i\in
V_2}\widetilde{\theta}_i(t_1)\leq
(1-\vartheta)A$, and for $k\geq
0$ and $i\in V$, $|\alpha_i(t_k)-\alpha_i(0)|\leq \mu$, where
$\alpha_i(t_k)=\frac{n_{i2}(t_k)}{n_{i1}(t_k)+n_{i2}(t_k)}$, then,
we have for $k\geq 1$, $\max_{i\in
V_1}\widetilde{\theta}_i(t_k)\leq \gamma^{k-1}A$, and $\max_{i\in
V_2}\widetilde{\theta}_i(t_k)\leq (1-\vartheta)\gamma^{k-1}A$,
where $\gamma=\max_{i\in V}(1-(\alpha_i(0)-\mu) \vartheta).$
\end{lem}

\textbf{Proof.} We prove the lemma by virtue of the mathematical
induction. First, the lemma holds for $k=1$. We assume that the
lemma holds at discrete-time instant $t_k$ with $k\geq1$, i.e., the
inequalities $\max_{i\in V_1}\widetilde{\theta}_i(t_k)\leq
 \gamma^{k-1}A$ and $\max_{i\in
V_2}\widetilde{\theta}_i(t_k)\leq (1-\vartheta)\gamma^{k-1}A$ hold.
Thus, for a follower agent $i\in V_1$, we have by \dref{4.5} \ban
&&\max_{i\in V_1}\widetilde{\theta}_i(t_{k+1})=\max_{i\in
V_1}\Big|\frac{\sum_{j\in\mathcal{N}_{i}(t_k)}\theta_j(t_k)}{n_{i}(t_k)}
-\overline{\theta}_0\Big|\nn\\&=&\max_{i\in
V_1}\bigg|\frac{\sum_{j\in\mathcal{N}_{i1}(t_k)}
(\theta_j(t)-\overline{\theta}_0)}{n_{i1}(t_k)+n_{i2}(t_k)}\nn\\&&\hskip 2cm
+\frac{\sum_{j\in
\mathcal{N}_{i2}(t_k)}(\theta_j(t_k)-\overline{\theta}_0)}{n_{i1}(t_k)+n_{i2}(t_k)}\bigg|
\nn\\&\leq& \max_{i\in V_1}\big\{(1-\alpha_i(t_k))\max_{i\in
V_1}\widetilde{\theta}_i(t_k)+\alpha_i(t_k)\max_{i\in
V_2}\widetilde{\theta}_i(t_k)\big\}\nn\\&\leq&\max_{i\in V_1}
\left\{(1-\alpha_i(t_k))\gamma^{k-1}
A+\alpha_i(t_k)(1-\vartheta)\gamma^{k-1} A
\right\}\nn\\&\leq&\max_{i\in V_1}(1-(\alpha_i(0)-\mu)\vartheta)
\gamma^{k-1}A\leq \gamma^{k}A.\ean While for a leader agent $i\in
V_2$, using \dref{4.2} we have \ban&& \max_{i\in
V_2}\widetilde{\theta}_i(t_{k+1})=(1-\vartheta)\max_{i\in
V_2}\Big|\frac{\sum_{j\in\mathcal{N}_{i1}(t_k)}
(\theta_j(t_k)-\overline{\theta}_0)}{n_{i1}(t_k)+n_{i2}(t_k)}\nn\\&&\hskip 1cm+\frac{\sum_{j\in
\mathcal{N}_{i2}(t_k)}
(\theta_j(t_k)-\overline{\theta}_0)}{n_{i1}(t_k)+n_{i2}(t_k)}\Big|
\leq (1-\vartheta)\nn\\&&\max_{i\in V_2}\Big\{
(1-\alpha_i(t_k))\max_{i\in
V_1}\widetilde{\theta}_i(t_k)+\alpha_i(t_k)\max_{i\in
V_2}\widetilde{\theta}_i(t)\Big\}\nn\\&&\leq(1-\vartheta)\gamma^{k}A.\ean
Thus, the results of the lemma holds at discrete-time instant
$t_{k+1}$. This completes the proof of the lemma. $\hfill\square$

For the moving speed, we have a similar result.
\begin{lem}\label{lambda}If there exist positive constants $B$ and
$ \mu$, such that $\max_{i\in V_1}\widetilde{v}_i(t_1)\leq B$,
$\max_{i\in V_2}\widetilde{v}_i(t_1)\leq
(1-\vartheta)B$ and for $k\geq
0,\q i\in V$, $|\alpha_i(t_k)-\alpha_i(0)|\leq \mu$, then for $k\geq 1$, we have $\max_{i\in
V_1}\widetilde{v}_i(t_k)\leq \gamma^{k-1}B$ and $\max_{i\in
V_2}\widetilde{v}_i(t_k)\leq (1-\vartheta)\gamma^{k-1}B.$
\end{lem}

The proof of Lemma  \ref{lambda} is similar to that of Lemma \ref{delta}, and we omit the proof details.

Similar to the analysis in Section \ref{leaderless}, we divide the
unit square $[0,1]^2$ into $M_n$ equally small squares labeled as
$1, 2,\cdots, M_n$, with $M_n=\lceil \frac{1}{a_n}\rceil^2$ where
$a_n$ satisfies $\sqrt{\log n/n}\ll a_n\ll1$. Denote $N_{k1}$ and
$N_{k2}$ with $1\leq k\leq M_n$  as the number of followers and
leaders in the corresponding small square, respectively. Introduce
the following sets \bna\label{square0} &&
B^{\prime}_n=\{\omega:\max_{1\leq j\leq M_n}N_{j1}=na_n^2(1+o(1)),\nn\\&&\hskip 2cm
\max_{1\leq j\leq M_n}N_{j2}=\rho_n
a_n^2(1+o(1)\},\\&&\label{theta00}\Theta^{\prime}_n=\big\{\omega:
\max_{i\in V}\big|\sum_{j\in
\mathcal{N}_i(0)}\theta_j(0)\big|=O(f_n)\big\},\\&&\label{v0}
\Gamma^{\prime}_n=\big\{\omega: \max_{i\in V}\big|\sum_{j\in
\mathcal{N}_i(0)}(v_j(0)-\frac{v_n}{2})\big|=O(v_n
f_n)\big\} ,\ena where $f_n=\sqrt{(n+\rho_n)r_n^2\log n}$. By
multi-array martingale lemma (Lemma 7 given by \citeasnoun{Auto}),  we have
$\mathbb{P}(B_n^{\prime})=1$ for large $n$ if $\alpha_n\gg
\frac{\log n}{nr_n^2}$. Similar to the synchronization analysis for
the leaderless model, and using the independency of the orientation
and heading at the initial time instant, we have
$\mathbb{P}(\Theta^{\prime}_n\bigcap\Gamma^{\prime}_n\bigcap
B^{\prime}_n)=1$. All of the following analysis is proceeded on the
set $\Theta^{\prime}_n\bigcap\Gamma^{\prime}_n\bigcap B^{\prime}_n$
without further explanations. Based on this, we can give some
estimations on the characteristics concerning the initial states,
which are presented in \ref{estiamtion}.


We characterize the change of the follower neighbors and leader
neighbors of a robot $i\ (i\in V) $ by the following two sets:
\bna\label{eq6}&&\mathcal{R}_{i1}=\{j\in V_1: (1-\eta)r_n\leq
\Delta_{ij}(0)\leq (1+\eta)r_n\},\q\q \\ \label{eq7}
&&\mathcal{R}_{i2}=\{j\in V_2: (1-\eta)r_n\leq \Delta_{ij}(0)\leq
(1+\eta)r_n\},\ena where the positive constant $\eta$ satisfies
$0<\eta\leq \frac{1}{512}$.  We denote the cardinality of the sets
$\mathcal{R}_{i1}$ and $\mathcal{R}_{i2}$ as $r_{i1}$ and $r_{i2}$,
respectively.

Here we briefly address why a certain number of leaders is needed in
order to guarantee that the followers track the reference signal of
leaders. Suppose that a very small number of leaders are added into
the system, for example, only one leader. Then the influence of the
leader is very weak, resulting in a low tracking rate. While the
leaders converge to the desired state with a certain rate. As a
consequence, all agents may form two disjoint clusters before their
orientations and speeds are synchronized to the desired states:
leader cluster and follower cluster, and it is impossible for the
followers to track the behavior of leaders.

Intuitively, the larger the neighborhood radius, the easier the
initial neighbor graph has a spanning tree; The smaller the moving
speed and dwell time, the easier the spanning tree is kept during
the evolution. For such a situation, the smaller the ratio of the
number of leaders will be needed to track the reference signal. In
the following theorem, we illustrate this intuition from a
theoretical point of view, and establish a quantitative result for
the ratio of the number of leaders needed.

\begin{thm}\label{largenumber1}Assume that the neighborhood
radius satisfies $\sqrt{\frac{\log n}{n}}\ll r_n\ll 1$. If the ratio
$\alpha_n$ of the number of leaders to the number of followers
satisfies one of the following two conditions:

\begin{enumerate}
  \item $ \vartheta \alpha_n\geq \frac{8v_n
\tau_n(1+|\overline{\theta}_0|)(1+o(1))}{\eta r_n}$, provided that
$v_n\tau_n\gg\frac{\log n}{n r_n}$;
  \item $ \alpha_n\gg \frac{\log n}{nr_n^2}$, provided that
$v_n\tau_n\ll\frac{\log n}{n r_n}$ or $v_n\tau_n=\Theta(\frac{\log
n}{n r_n})$,
\end{enumerate}
then all robots move with $\overline{\theta}_0$ and $v_n$
eventually.
\end{thm}

\textbf{Proof.}  Lemmas \ref{delta} and \ref{lambda} show that
the estimation of $|\alpha_i(t_k)-\alpha_i(0)|$ is a key step for
the convergence of $\max_{i\in V}\widetilde{\theta}_i(t_k)$ and
$\max_{i\in V}\widetilde{v}_i(t_k)$.  $\alpha_i(t)$ is defined via
the number of leader neighbors and the number of follower neighbors
of the agent $i$ at time $t$. In order to estimate
$|\alpha_i(t_k)-\alpha_i(0)|$, we show that the distance between any
pair of robots $i$ and $j$ at any discrete-time instant $t_k$
satisfies the following inequality, \bna\label{eq8}
|\Delta_{ij}(t_k)-\Delta_{ij}(t_0)|\leq \eta r_n, k\geq 0,\ena where
$\eta$ is a positive constant taking the same value as that in
\dref{eq6} and \dref{eq7}. If \dref{eq8} holds, then for a robot
$i$, the change of its follower neighbors and leader neighbors at
time $t_k$ in comparison with those at the initial time is included
in the sets $\mathcal{R}_{i1}$ and $\mathcal{R}_{i2}$ defined by
\dref{eq6} and \dref{eq7}, respectively. For $i\in V$, we have
$|n_{i1}(t_{k})-n_{i1}(0)|\leq \max_{i\in V}r_{i1}$ and
$|n_{i2}(t_{k})-n_{i2}(0)|\leq \max_{i\in V}r_{i2}.$ By the
estimation of the number of follower neighbors and leader neighbors
at the initial time instant given in \ref{estiamtion}, we
have \bna\label{6.4} && \max_{i\in
V}\left|\alpha_i(t_{k})-\alpha_i(0)\right|\nn\\&=&\max_{i\in
V}\left|\frac{n_{i2}(t_{k})}{n_{i1}(t_{k})+n_{i2}(t_{k})}-
\frac{n_{i2}(0)}{n_{i1}(0)+n_{i2}(0)}\right|\nn\\&\leq&
\frac{\dmax_{i\in V}n_{i2}(0) \dmax_{i\in V}r_{i1}+\dmax_{ i\in
V}n_{i1}(0)\dmax_{i\in V}r_{i2}}{\dmin_{ i\in
V}\{(n_{i1}(t_{k})+n_{i2}(t_{k}))(n_{i1}(0)+n_{i2}(0))\}}\nn\\&\leq&
\frac{8\eta \alpha_n(n\pi r_n^2)^2(1+o(1))}{\frac{(n\pi
r_n^2)^2}{16}(1-16\eta)(1+\alpha_n)^2}\leq 256 \eta
\alpha_n\leq\frac{\alpha_n}{2}.\ena Using Lemmas \ref{delta} and
\ref{lambda}, as $k\rightarrow\infty$,  $\max_{i\in
V}\widetilde{\theta}_i(t_k)\leq
(1-\frac{\vartheta\alpha_n(1+o(1))}{2})^{k-1}
(|\overline{\theta}_0|+L_n)\rightarrow0$  and $\max_{i\in V}\widetilde{v}_i(t_k)\leq
(1-\frac{\vartheta\alpha_n(1+o(1))}{2})^{k-1}\frac{v_n}{2}(1+o(1))\rightarrow0$. Moreover, by \dref{10.1} and
\dref{4.4}, as $t\rightarrow\infty$, $\max_{i\in V_1}\widetilde{\theta}_i(t)\leq \max_{i\in
V}\widetilde{\theta}_i(t_k)\rightarrow0
$ and $\max_{i\in
V_2}\widetilde{\theta}_i(t)\leq (1-\frac{\vartheta(t-t_k)}{\tau_n})
\max_{i\in V}\widetilde{\theta}_i(t_k)\rightarrow0$,  which mean that both the followers and
leaders will move with the same desired orientation
$\overline{\theta}_0$ eventually. By a similar analysis, we can prove that
all robots move with the same desired speed $v_n$
eventually.

Now, we use the mathematical induction to prove \dref{eq8}. It is
clear that \dref{eq8} holds for $k=0$. We assume that it holds for
$0\leq l\leq k_0$. By the analysis of \dref{6.4}, we have
$|\alpha_i(t_l)-\alpha_i(0)|\leq \frac{\alpha_n}{2}$. Using Lemmas
\ref{delta} and \ref{lambda}, for $0\leq l\leq k_0$, we have $ \max_{i\in
V_1}|\widetilde{\theta}_i(t_{l+1})|\leq
\left(\widehat{\alpha}_n\right)^{l}(|\theta_0|+L_n)$, $
\max_{i\in V_2}|\widetilde{\theta}_i(t_{l+1})|\leq(1-\vartheta)
\left(\widehat{\alpha}_n\right)^{l}(|\theta_0|+L_n)$, and $
\max_{i\in V_1}|\widetilde{v}_i(t_{l+1})|\leq
\left(\widehat{\alpha}_n\right)^{l}\frac{v_n(1+o(1))}{2}$,
$\max_{i\in V_2}|\widetilde{v}_i(t_{l+1})| $ $\leq  (1-\vartheta)
\left(\widehat{\alpha}_n\right)^{l}\frac{v_n(1+o(1))}{2}$,
where $L_n$ is defined in Lemma \ref{thetaleader} in
\ref{estiamtion}, and
$\widehat{\alpha}_n=\left(1-\frac{\vartheta\alpha_n}{2}(1+o(1))\right)$.
For followers, using \dref{1.4} and \dref{2.4.1}, we have for $t\in
[t_l, t_{l+1}], l=0,1,2,\cdots, k_0$, \ban &&\max_{i\in
V_1}|\widetilde{\theta}_i(t)|\leq
\left(1-\frac{t-t_l}{\tau_n}\right)\max_{i\in
V_1}|\widetilde{\theta}_i(t_l)|\nn\\&&\hskip 2cm+\max_{i\in
V_1}\left|\frac{t-t_l}{\tau_n n_i(t_l)}\sum_{j\in
\mathcal{N}_i(t_{l})}\widetilde{\theta}_j(t_l)\right|\nn\\&\leq&\left(1-\frac{t-t_l}{\tau_n}\right)\max_{i\in
V_1}|\widetilde{\theta}_i(t_l)|+\nn\\&&\frac{t-t_l}{\tau_n}\max_{i\in
V_1}((1-\alpha_i(t_l))\max_{i\in
V_1}|\widetilde{\theta}_i(t_l)|+\alpha_i(t_l)\max_{i\in
V_2}|\widetilde{\theta}_i(t_l)| \nn\\&\leq&
\left(1-\frac{\alpha_n\vartheta(t-t_l)(1+o(1))}{2\tau_n}\right)\left(\widehat{\alpha}_n\right)^{l-1}(|\theta_0|+L_n).\ean
Similarly, $\max_{i\in
V_1}|\widetilde{v}_i(t)|\leq(1-\frac{\alpha_n\vartheta(t-t_l)(1+o(1))}{2\tau_n})(\widehat{\alpha}_n)^{l-1}$ $\cdot\frac{v_n(1+o(1))}{2}.$

While for leaders, using \dref{4.4} and \dref{2.4}, we have $\dmax_{i\in V_2}|\widetilde{\theta}_i(t)|\leq (1-\vartheta)
(1-\frac{\alpha_n\vartheta(t-t_l)(1+o(1))}{2\tau_n})(\widehat{\alpha}_n)^{l-1}(|\theta_0|+L_n)$, and $ \dmax_{i\in
V_2}|\widetilde{v}_i(t)|\leq(1-\vartheta)(1-\frac{\alpha_n\vartheta(t-t_k)(1+o(1))}{2\tau_n})(\widehat{\alpha}_n)^{l-1}\frac{v_n(1+o(1))}{2}.$

Using Lemma \ref{thetaleader} in  \ref{estiamtion}, the
following result can be obtained, \bna\label{3.2}&&
\sum_{l=1}^{k_0}\max_{i\in V}v_i(t_l)\int_{t_l}^{t_{l+1}}\max_{i\in
V}|\widetilde{\theta}_i(t)|dt\leq v_n\sum_{l=1}^{k_0}\int_{t_l}^{t_{l+1}}\nn\\&&
\big\{(1-\frac{\alpha_n\vartheta(t-t_l)(1+o(1))}{2\tau_n})(\widehat{\alpha}_n)^{l-1}(|\theta_0|+L_n)\big\}dt
\nn\\&=&\sum_{l=1}^{k_0}\frac{(1+\widehat{\alpha}_n)}{2}v_n\tau_n\widehat{\alpha}_n^{l-1}(|\theta_0|+L_n)(1+o(1))\nn\\&=&
\frac{(1+\widehat{\alpha}_n)v_n\tau_n(|\theta_0|+L_n)}{\vartheta\alpha_n}(1+o(1)).\ena
Moreover, we have \bna\label{3.2.1} &&
\sum_{l=1}^{k_0}\int_{t_l}^{t_{l+1}}\max_{i\in V}
|\widetilde{v}_i(t)|dt\leq
v_n\sum_{l=1}^{k_0}\int_{t_l}^{t_{l+1}}\nn\\&&\big\{(1-\frac{\alpha_n\vartheta(t-t_k)(1+o(1))}{2\tau_n})
(\widehat{\alpha}_n)^{l-1}(1+o(1))\big\}dt\nn\\
&\leq&\sum_{l=1}^{k_0}\frac{(1+\widehat{\alpha}_n)v_n\tau_n\widehat{\alpha}_n^{l-1}(1+o(1))}{2}\nn\\&=&
\frac{(1+\widehat{\alpha}_n)v_n\tau_n(1+o(1))}{\vartheta\alpha_n}.
\ena
Thus, using \dref{3.2} and \dref{3.2.1}, we obtain \bna\label{5.3}&&
|\Delta_{ij}(t_{k_0+1})-\Delta_{ij}(0)|\leq\nn\\&&
\sum_{l=0}^{k_0}\|X_i(l+1)-X_{j}(l+1)-X_i(l)+X_{j}(l)\|_2 \nn\\
&\leq&
4v_n\tau_n+4\sum_{l=1}^{k_0}\Big\{\int_{t_l}^{t_{l+1}}\max_{i\in
V}|\widetilde{v}_i(t)|dt\nn\\&& +\max_{i\in
V}v_i(t_l)\int_{t_l}^{t_{l+1}}\max_{i\in
V}|\widetilde{\theta}_i(t)|dt\Big\} \nn\\&\leq&
\frac{8v_n\tau_n(1+|\theta_0|)(1+o(1))}{\vartheta\alpha_n}\leq \eta
r_n,\ena  where the condition on the ratio of the
number of leaders is used in the last inequality. This completes
the proof of \dref{eq8}.\q$\square$


%

\subsection{ Dynamic tracking of unicycle robots}\label{dynamic}

For some complicated tasks, such as path following, avoiding
collision with obstacles, the reference signal of leaders may vary
with time. In this part, we consider the dynamic tracking of
unicycle robots. In order to present the problem and the result
clearly, we consider the case where the desired orientations of
leaders may change over time, but the desired speed keeps unchanged.
For a leader robot $i\ (i\in V_2)$, it has the following information
at discrete-time instant $t_k$, $\{v_j(t_k)-v_i(t_k),\ \theta_j(t_k)-\theta_i(t_k), \
\overline{\theta}_{k}-\theta_i(t_k),  v_{n}-v_i(t), \ j\in
\mathcal{N}_i(t_k)\}$,  where $v_n$ is the desired speed
and $\overline{\theta}_{k}$ is the desired orientation at time
$t_k$.

The notations, including the set of follower neighbors
$\mathcal{N}_{i1}(t)$, the set of leader neighbors
$\mathcal{N}_{i2}(t)$, $\mathcal{R}_{i1}$ and $\mathcal{R}_{i2}$
have the same meanings as those in Subsection \ref{problem}.


The closed-loop dynamics of followers is still described by
\dref{3.3}. For the leaders, we adopt the similar distributed
control law as that of \dref{3.6} for $t\in [t_k, t_{k+1}) (k=0, 1,
\cdots,)$, just replacing $\overline{\theta}_{0}$ by
$\overline{\theta}_{k}$. Thus, we obtain the closed-loop
dynamics of the orientation and speed of leaders for $t\in [t_k, t_{k+1})$,
 \bna &&\dot{\theta}_i(t)=\frac{1}{\tau_n}
\Big\{\vartheta(\overline{\theta}_{k}-\theta_i(t_k))\nn\\&&\hskip 2cm+\frac{1-\vartheta}{
d_i(t_k)}\dsum_{j\in\mathcal{N}_i(t_k)}(\theta_j(t_k)-\theta_i(t_k))\Big\},\nn\\ &&
\dot{v}_i(t)=
\frac{1}{\tau_n}\Big\{\vartheta(v_{n}-v_i(t_k))\nn\\&&\hskip 2cm+\frac{1-\vartheta}{
d_i(t_k)}\dsum_{j\in\mathcal{N}_i(t_k)}(v_j(t_k)-v_i(t_k))\Big\},
\ena where $\vartheta\leq1$ is a positive constant.

\begin{figure}[htbp]
\centering
    \includegraphics[width=8cm]{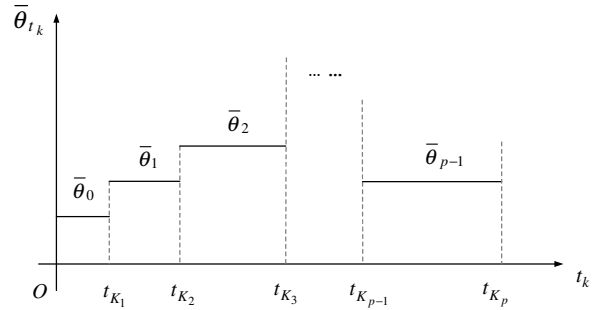}
  \caption{ The dynamic signal of the desired orientations.}
   \label{fgg1}
\end{figure}

For a large crowd, it is apparent both mathematically and
intuitively that if the desired orientation of leaders changes too
fast, then it is impossible for the followers to track. In this
paper, we consider the situation where the desired orientation of
the leaders is piecewise constant in such a way that it keeps
constant until the followers track the reference signal in the sense
that the maximum dissimilarity for the orientations is less than a
pre-defined tracking error $\varepsilon>0$, as shown in Fig.
\ref{fgg1} where the time instants $t_{K_i}$ depend on the tracking
error. Although this assumption seems to be rather restrictive, it
is nevertheless reasonable for applications such as crowd control by
active intervention.

Denote the difference between the desired orientations at two
contiguous time instants as
$D_k=|\overline{\theta}_{k+1}-\overline{\theta}_{k}|$. We assume
$\sum_{k=0}^\infty D_k<\infty$. Since what we concern is the
tracking effect of the robots at each stage $[T_{k_{l-1}}, T_{k_l}],
l=1, 2, \cdots$, we need to analyze the dynamical behavior of the
leaders and followers stage by stage, and the ending states of the
robots at the latest stage is just the starting states at the
current stage. We give the main result for the system with dynamic
leaders as follows.

\begin{thm}\label{dynamictrack} Assume that the neighborhood radius satisfies $\sqrt{\frac{\log n}{n}}\ll r_n\ll 1$.
Then for any given tracking error $\varepsilon>0$, we have for $i\in
V$, $\lim_{t\rightarrow\infty}\left|v_i(t)-v_n\right|=0$ and
\bna\label{or} \max_{i\in
V}\left|\theta_i(t_{K_{l+1}})-\overline{\theta}_{{l}}\right|\leq
\varepsilon, \ \ l=0, 1, 2, \cdots,
 \ena if the
ratio of the number of leaders satisfies one of the following
conditions

1) $ \vartheta \alpha_n\geq \frac{4v_n
\tau_n(1+\sum_{k=0}^{\infty}D_k+|\overline{\theta}_0|)(1+o(1))}{\eta
r_n}$ provided that $v_n\tau_n\gg\frac{\log n}{n r_n}$.

2) $ \alpha_n\gg \frac{\log n}{nr_n^2}$ provided that
$v_n\tau_n\ll\frac{\log n}{n r_n}$ or $v_n\tau_n=\Theta(\frac{\log
n}{n r_n})$.
\end{thm}

\textbf{Proof.}  We first prove that for any pair of robots $i$ and $j$, we
have \bna
\label{6.6.1}&&\left|\Delta_{ij}(t_k)-\Delta_{ij}(0)\right|\leq \eta
r_n,\q \forall k\geq 0. \ena Using Theorem \ref{largenumber1}, we
see that \dref{6.6.1} holds for $t\in [0, t_{K_1}]$ by taking
$K_1\geq\Big\lceil\frac{\log
(\varepsilon/(|\overline{\theta}_0|+L_n)}{\log
\widehat{\alpha}_n}\Big\rceil+1$. For $0\leq l\leq K_1$, we have
$\max_{i\in
V}|\alpha_i(t_l)-\alpha_i(0)|\leq\frac{\alpha_n}{2}(1+o(1))$. Thus,
$\max_{i\in
V_1}|\theta_i(t_{K_1})-\overline{\theta}_0|\leq
\widehat{\alpha}_n^{K_1-1} (|\overline{\theta}_0|+L_n)\leq
\varepsilon$, and $\max_{i\in
V_2}|\theta_i(t_{K_1})-\overline{\theta}_0|\leq
(1-\vartheta)\widehat{\alpha}_n^{K_1-1} (|\overline{\theta}_0|+L_n)\leq (1-\vartheta)\varepsilon$, 
where
$\widehat{\alpha}_n=\left(1-\frac{\vartheta\alpha_n(1+o(1))}{2}\right)$.

We now analyze the dynamical behavior of all agents for $k\in[K_p+1,
K_{p+1}]$ $(p=2, 3, \cdots)$ with $
K_p=\Big\lceil\frac{\log (\varepsilon/(\varepsilon+D_{p-1}))}{\log
\widehat{\alpha}_n}\Big\rceil+\sum_{s=1}^{p-1}K_s+1.$ We have
at the time instant $t_{K_p}$, $ \max_{i\in
V_1}|\theta_i(t_{K_p})-\overline{\theta}_{p-1}|\leq \varepsilon$, and
$\max_{i\in V_2}|\theta_i(t_{K_p})-\overline{\theta}_{p-1}|\leq
(1-\vartheta)\varepsilon.$ At the time instant $t_{K_p+1}$, the
desired orientation of leaders changes to $\overline{\theta}_p$, and
the orientations of the followers and leaders at time $t_{K_p+1}$
satisfy $\max_{i\in
V_1}|\theta_i(t_{K_p+1})-\overline{\theta}_{p}|\leq
(1-\alpha_i(t_{K_p}))\Big\{\max_{i\in
V_1}|\theta_i(t_{K_p})-\overline{\theta}_{p-1}|+D_{p-1}\Big\}+\alpha_i(t_{K_p})\Big\{\max_{i\in
V_2}|\theta_i(t_{K_p})-\overline{\theta}_{p-1}|+D_{p-1}\Big\}\leq\varepsilon+D_{p-1}$, and $\max_{i\in V_2}|\theta_i(t_{K_p+1})-\overline{\theta}_p|\leq
(1-\vartheta)(\varepsilon+D_{p-1})$, where
$D_{p-1}=|\overline{\theta}_p-\overline{\theta}_{p-1}|$. Assume that
\dref{6.6.1} holds for $k_0\in [K_p+1, K_{p+1}], p=2, 3, \cdots $.
Then we have $\max_{i\in
V}|\alpha_i(t_l)-\alpha_i(0)|\leq\frac{\alpha_n}{2}(1+o(1))$ for $
l\leq k_0$ . Using Lemmas \ref{delta} and \ref{lambda}, we have for
$k\in [K_p+1, k_0], p=2,\cdots, P-1$, $ \max_{i\in
V_1}|\theta_i(t_{k})-\overline{\theta}_p|\leq
\widehat{\alpha}_n^{k-K_{p}-1} (\varepsilon+D_{p-1})$, and
$\max_{i\in V_2}|\theta_i(t_{k})-\overline{\theta}_p|\leq
(1-\vartheta)\widehat{\alpha}_n^{k-K_{p}-1}
(\varepsilon+D_{p-1}).$  Thus, the distance between agents $i$
and $j$ satisfies
 \bna\label{5.3.3} &&
|\Delta_{ij}(t_{k_0+1})-\Delta_{ij}(0)|\leq\Big(\sum_{s=0}^{p-1}\sum_{l=K_{s}+1}^{K_{s+1}}+\sum_{l=K_{p}+1}^{k_0}\Big)\nn\\&&
\Big\{\Big|\int_{t_l}^{t_{l+1}}\left[v_i(t)\cos\theta_i(t)-v_j(t)\cos\theta_j(t)\right]dt\Big|
\nn\\ &&
+\Big|\int_{t_l}^{t_{l+1}}\left[v_i(t)\sin\theta_i(t)-v_j(t)\sin\theta_j(t)\right]dt\Big|\Big\}\nn\\&\leq&
2(1+\widehat{\alpha}_n)v_n\tau_n(1+o(1))\Big(\sum_{l=0}^{K_1}\widehat{\alpha}_n^{l-1}(|\overline{\theta}_0|+L_n)\nn\\&&+\sum_{s=1}^{p-1}
\sum_{k=K_s+1}^{K_{s+1}}\widehat{\alpha}_n^{l-K_s-1}(\varepsilon+D_{s-1})
\nn\\&&+\sum_{l=K_p+1}^{k_0}\widehat{\alpha}_n^{l-K_p-1}(\varepsilon+D_{p-1})
\Big)\nn\\&&+\sum_{l=0}^{k_0}
(1+\widehat{\alpha}_n)v_n\tau_n\widehat{\alpha}_n^{l-1}(1+o(1))\nn\\
&\leq&8v_n\tau_n\frac{1}{\vartheta\alpha_n}(1+o(1))\left(|\overline{\theta}_0|+\sum_{s=0}^{p-1}D_s+1\right)\leq
\eta r_n.\ena

By the above analysis, we see that for $k\in[K_p+1, K_{p+1}], p=0,
1, \cdots,$  the assertion
\dref{or} holds, and we have  for $l\geq 0$, $\max_{i\in
V}|\alpha_i(t_l)-\alpha_i(0)|\leq\frac{\alpha_n}{2}(1+o(1)).$
Moreover, using Lemma \ref{lambda}, we obtain \ban &&\max_{i\in
V_1}\widetilde{v}_i(t_k)\leq
(\widehat{\alpha}_n)^{k-1}\frac{v_n(1+o(1))}{2}\rightarrow0,\ \
\hbox{as}\ \ k\rightarrow\infty,\nn\\ && \max_{i\in
V_2}\widetilde{v}_i(t_k)\leq
(1-\vartheta)(\widehat{\alpha}_n)^{k-1}\frac{v_n(1+o(1))}{2}\rightarrow0,
\ \ \hbox{as}\ \  k\rightarrow\infty.\ean \q$\square$

\begin{rem} A direct consequence is that the ratio of the number of leaders to the number of
followers depends on the times and the amplitude that the desired
orientations change during evolution. \end{rem}

\subsection{A simulation example}
\begin{figure}[htbp]
\centering
    \includegraphics[width=8cm]{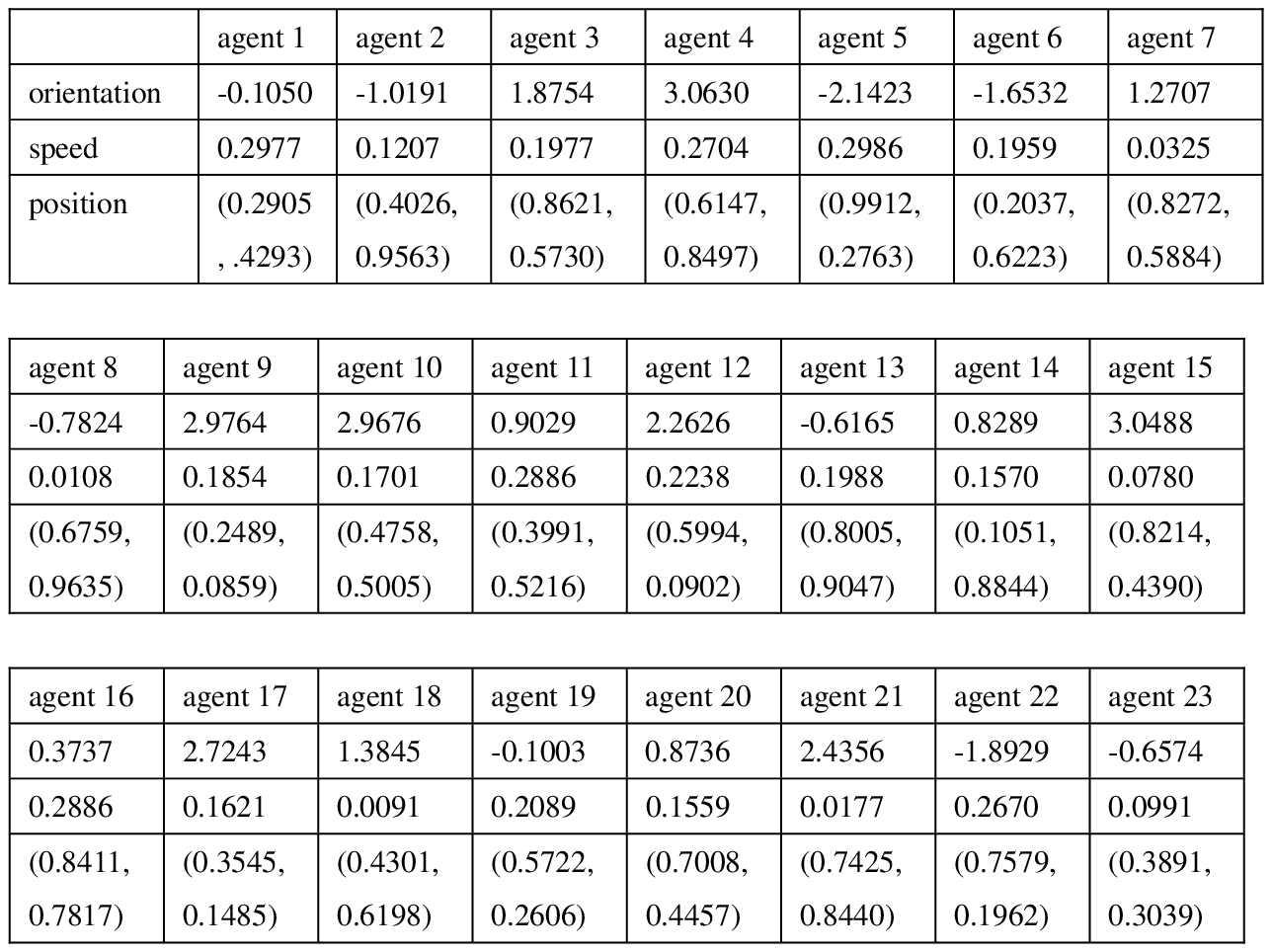}
  \caption{ The initial states of all agents in the simulation example.}
   \label{fgg2}
\end{figure}

\begin{figure}[htbp]
\centering
 \includegraphics[width=7cm]{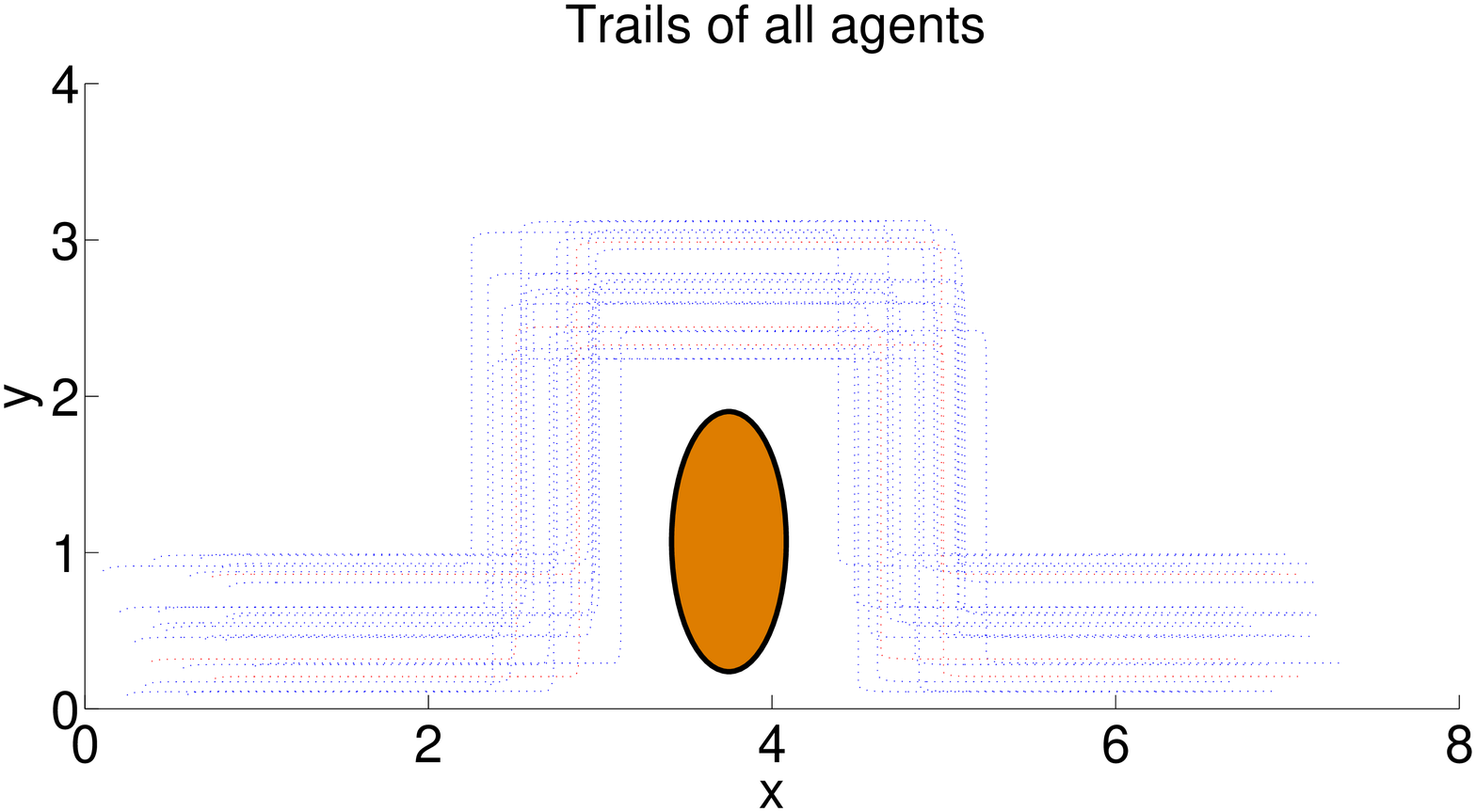}
\includegraphics[width=7cm]{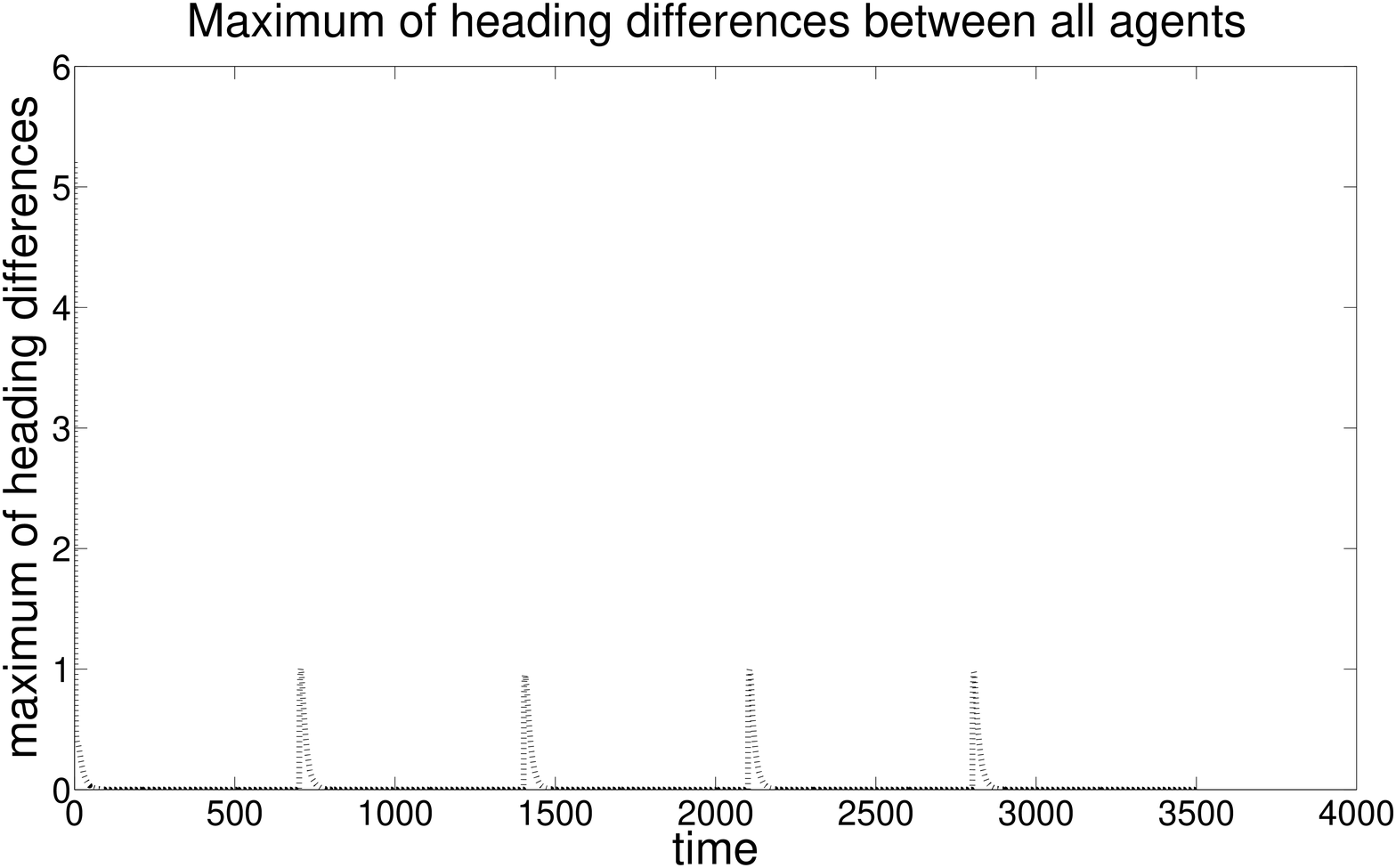}
\includegraphics[width=7cm]{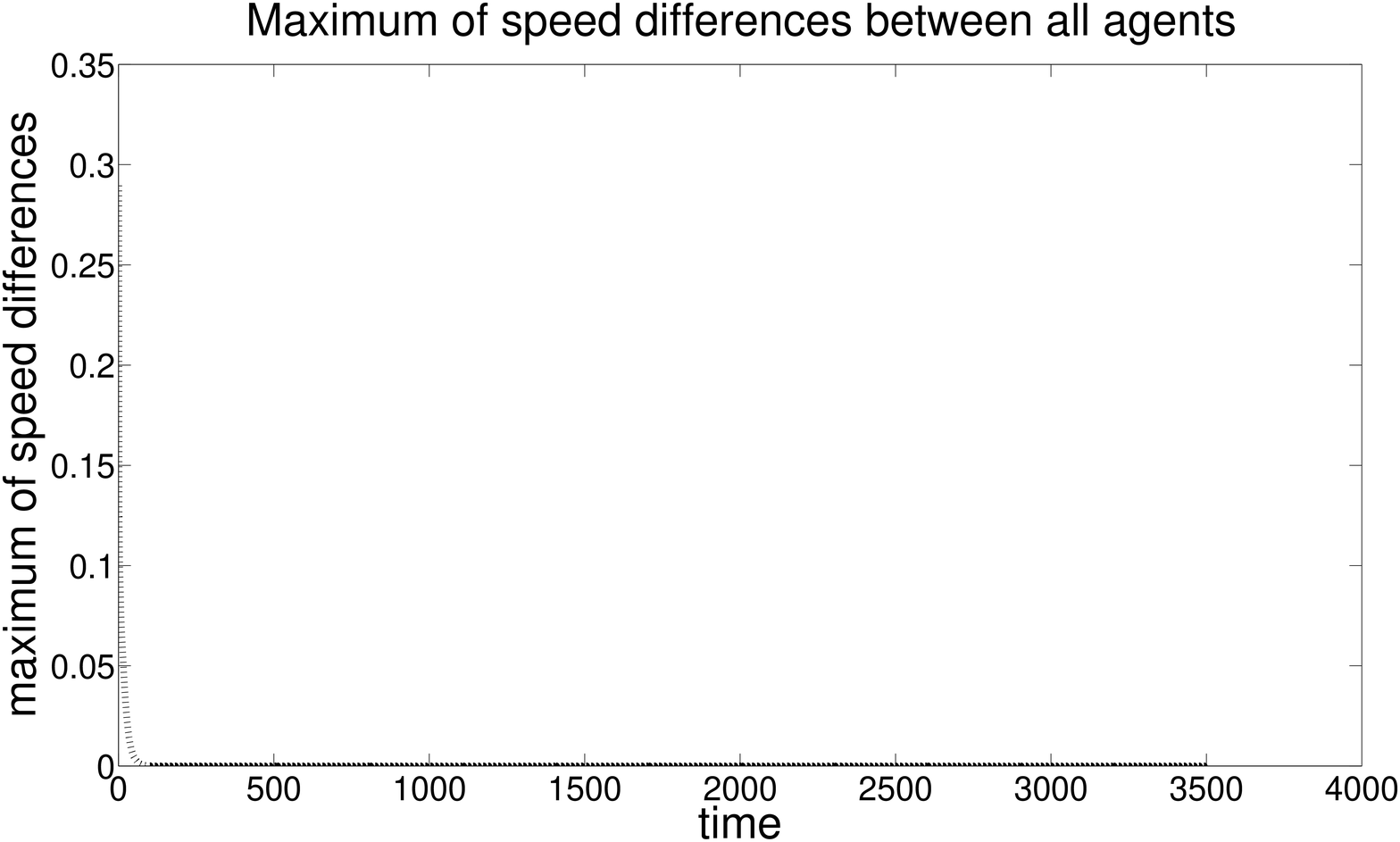}   
\caption{The trajectories of all robots, where the red
and blue lines, respectively, denote the trajectories of the leaders
and followers.}  
   \label{fgg3}
\end{figure}

We illustrate the feasibility of guiding a group of ordinary robots
to accomplish a complicated task by introducing dynamic leaders
whose desired orientation may change with the requirement of the
task. The system is composed of $20$ ordinary agents labeled $1, 2,
\cdots, 20$, and we aim to guide these robots to move to the right
along the bottom line, go across an oval-shaped obstacle and not to
collide with it as shown in Fig. \ref{fgg3}. The desired orientations
can be taken as $\overline{\theta}_0=0$,
$\overline{\theta}_1=\frac{\pi}{2}$, $\overline{\theta}_2=0$
$\overline{\theta}_3=-\frac{\pi}{2}$, $\overline{\theta}_4=0$. To
complete such a task, we introduce 3 leaders into the system which
are labeled $21, 22$ and $23$. The initial states of all agents are
taken to satisfy Assumption \ref{initial} with the maximum initial
speed $v=0.3$, and the neighborhood radius $r=0.3$. The dwell time
is taken as $\tau=0.01$. The initial orientations, velocities and
positions of the leaders and followers are listed in Fig. \ref{fgg2}.
Fig. \ref{fgg3} shows the trajectories of all robots, and we see that
the leaders can guide the followers to achieve the pre-defined task.

\section{Concluding Remarks} \label{concluding}
In this paper, we proposed a sampled-data distributed control law
for a group of nonholonomic unicycle robots, and established
sufficient conditions for synchronization for the leaderless case
without relying on dynamical properties of the neighbor graphs. In
order to steer the system to a desired state, we then introduced
leaders with constant or time-varying signals into the system, and
provided the proportion of the leaders needed to track the static or
dynamic signals. In our model, the robots are connected via
distance-induced graphs, and a dwell time is assumed for the
feasibility of information sensing and processing which as a
consequence  avoids issues such as chattering caused by abrupt
changes of the neighbor relations. Some interesting problems deserve
to be further investigated, for example, how to design the control
law based on the information of relative orientations and relative
positions, how to design the sampled-data control law to avoid
collisions, and how to design the observer-based control laws for the case where the relative speed and relative orientation can not be directly measured.
\appendix
\def\thesection      {\appendixname\ \Alph{section}.}
\def\theequation     {\Alph{section}.\arabic{equation}}

\section{Proof of the inequalities \dref{distancerecur} and \dref{average}.}\label{prooflemma}

\textbf{Proof of \dref{distancerecur}.} By \dref{2.3}, we have $
x_i(t_{k_0+1})-x_i(t_{k_0})=\int_{{t_{k_0}}}^{t_{{k_0}+1}}v_i(t)\cos\theta_i(t)dt$
and
$y_i(t_{{k_0}+1})-y_i(t_{k_0})=\int_{t_{k_0}}^{t_{{k_0}+1}}v_i(t)$ $\sin\theta_i(t)dt$.
Denote $X_i(t_k)=(x_i(t_k), y_i(t_k))^{\prime}$. The distance
between any two agents $i$ and $j$ satisfies the following
inequality: \bna\label{distant}&&
|\Delta_{ij}(t_{k+1})-\Delta_{ij}(t_k)|\nn\\&=&\big|\|X_i(t_{k+1})-X_j(t_{k+1})\|-\|X_i(t_k)-X_j(t_k)\|\big|\nn\\&\leq&
\|X_i(t_{k+1})-X_i(t_k)-(X_{j}(t_{k+1})-X_j(t_k))\| \nn\\&=&
\left\|\left(
                     \begin{array}{c}
                       \int_{t_k}^{t_{k+1}}[v_i(t)\cos\theta_i(t)-v_j(t)\cos\theta_j(t)]dt \\
                       \int_{t_k}^{t_{k+1}}[v_i(t)\sin\theta_i(t)-v_j(t)\sin\theta_j(t)]dt \\
                     \end{array}
                   \right)
\right\|\nn\\ &\leq&
\int_{t_k}^{t_{k+1}}\Big\{\left|v_i(t)\cos\theta_i(t)-v_j(t)\cos\theta_j(t)\right|\nn\\&&+
\left|v_i(t)\sin\theta_i(t)-v_j(t)\sin\theta_j(t)\right|\Big\}dt.\ena
Using \dref{3.9}, the first term of \dref{distant} satisfies
\bna\label{b.2} &&
\int_{t_k}^{t_{k+1}}\left|v_i(t)\cos\theta_i(t)-v_j(t)\cos\theta_j(t)\right|dt\nn\\&\leq&
\int_{t_k}^{t_{k+1}}|(v_i(t)-v_j(t))\cos\theta_i(t)|\nn\\&&+|v_j(t)(\cos\theta_i(t)-\cos\theta_j(t))|dt\leq\int_{t_k}^{t_{k+1}}\delta v(t)dt\nn\\&&
+2\max_{1\leq i\leq
n}v_i(t_k)\int_{t_k}^{t_{k+1}}|\sin\frac{\theta_i(t)-\theta_j(t)}{2}|dt\nn\\&\leq&\int_{t_k}^{t_{k+1}}\delta
v(t)dt+\max_{1\leq i\leq
n}v_i(t_k)\int_{t_k}^{t_{k+1}}\delta\theta(t)dt,\ena and the second
term of \dref{distant} satisifies \bna\label{b.3} &&
\int_{t_k}^{t_{k+1}}\left|v_i(t)\sin\theta_i(t)-v_j(t)\sin\theta_j(t)\right|dt\leq\int_{t_k}^{t_{k+1}}\delta v(t)dt\nn\\&&
+2\max_{1\leq i\leq
n}v_i(t_k)\int_{t_k}^{t_{k+1}}|\sin\frac{\theta_i(t)-\theta_j(t)}{2}|dt\nn\\&\leq&\int_{t_k}^{t_{k+1}}\delta
v(t)dt+\max_{1\leq i\leq
n}v_i(t_k)\int_{t_k}^{t_{k+1}}\delta\theta(t)dt.\ena Substituting
\dref{b.2} and \dref{b.3} into \dref{distant} yields the inequality
\dref{distancerecur}.\q$\square$

\textbf{Proof of \dref{average}.} We see that if at the initial time
instant, the distance between $i$ and $j$ satisfies
$\Delta_{ij}(t_0)< (1-\eta_n)r_n$, then by \dref{1.5} we have
$\Delta_{ij}(t_k)<r_n$; Otherwise, if $\Delta_{ij}(t_0)\geq
(1+\eta_n)r_n$, then by \dref{1.5} we have $\Delta_{ij}(t_k)\geq
r_n$. Compared with the initial time instants, the change of the
agent $i$'s neighbors at time instant $t_k$ is characterized by the
following set, \bna\label{1.1} \mathcal{R}_i=\left\{j: (1-\eta_n)
r_n\leq \Delta_{ij}(0)\leq (1+\eta_n) r_n\right\}.\ena

Denote the maximum number of agents in the set $\mathcal{R}_j$
defined by \dref{1.1} as $R_{\max}$.  By the fact $P(B_n)=1$ with
$B_n$ defined in \dref{square}, we have for large $n$, $R_{\max}
\leq  4n\pi\eta_n r_n^2(1+o(1))$. Since the inequality \dref{1.5}
holds for $l\leq k_0$, the number of each agent's neighbors changed
at time $t_{l+1}$ in comparison with its initial neighbors is
bounded by $R_{\max}$. Using Lemma 3 in \cite{tang2}, we have for
large $n$ \ban &&\|P(t_{l+1})-P(0))\|\leq
\frac{R_{\max}}{d_{\min}(0)}\cdot\frac{d_{\max}(0)-d_{\min}(0)}{d_{\min}(0)-R_{\max}}\nn\\&\leq&\frac{4n\pi\eta_n
r_n^2}{\frac{1}{4}n\pi r_n^2}\cdot\frac{n\pi r_n^2+\frac{1}{4}n\pi
r_n^2}{\frac{1}{4}n\pi r_n^2-4n\pi\eta_n
r_n^2}(1+o(1)=80\eta_n(1+o(1)).\ean
\q$\square$

\section{Estimation of some characteristics for the leader-follower model.}\label{estiamtion}

By  the fact $P(B_n^{\prime})=1$ with $B_n^{\prime}$ defined in
\dref{square0}, we can directly obtain the following results.
\begin{lem}\label{dddd}  Let the neighborhood radius $r_n$  and the ratio  $\alpha_n$ satisfy the conditions: $\sqrt{\log n/n}\ll
r_{n}\ll1$ and $\alpha_n\gg\frac{\log n}{nr_n^2}$.  Then the
following results hold almost surely for large $n$

1) For any agent $i\in V$, we have \ban
\alpha_i(0)=\frac{\alpha_n}{1+\alpha_n}(1+o(1))=\alpha_n(1+o(1)).\ean

2) The cardinality of the sets $\mathcal{N}_{i1}(0)$ and
$\mathcal{N}_{i2}(0)$ satisfy \ban &&\min_{i\in V}d_{i1}(0)=
\frac{n\pi r_n^2}{4}(1+o(1)),\nn\\&& \max_{i\in V}d_{i1}(0)= n\pi
r_n^2(1+o(1)); \nn\\&& \min_{i\in V}d_{i2}(0)=\frac{n\pi
r_n^2\alpha_n}{4}(1+o(1)), \nn\\&&\max_{i\in V}d_{i2}(0)= n\pi
r_n^2\alpha_n(1+o(1)).\ean

3) The number of agents in the sets $\mathcal{R}_{i1}$ and
$\mathcal{R}_{i2}$ satisfies  \ban&&
 \max_{i\in V}r_i=4\eta\pi n
r_n^{2}(1+o(1)),\nn\\&& \max_{i\in V}r_i^{\prime}=4\eta\pi
n\alpha_nr_n^{2}(1+o(1)).\ean
\end{lem}


\begin{lem}\label{thetaleader} Under the assumptions in Lemma \ref{dddd}, the following assertions hold almost surely for large $n$
\ban&& \max_{i\in V_1}|\widetilde{\theta}_i(t_1)|\leq
|\theta_0|+L_n,\nn\\&&  \max_{i\in V_2}|\widetilde{\theta}_i(t_1)|\leq
(1-w)(|\theta_0|+L_n),\\&& \max_{i\in
V_1}|\widetilde{v}_i(t_1)|=\frac{v_n}{2}(1+o(1)),\nn\\&& \max_{i\in
V_2}|\widetilde{v}_i(t_1)|=(1-w)\frac{v_n}{2}(1+o(1)), \ean where
$L_n=\frac{4C_1(1+o(1))}{\pi}\sqrt{\frac{\log n}{n r_n^2}}$.
\end{lem}

\bibliographystyle{dcu}        

\begin{thebibliography}{xx}

\harvarditem{ Cort$\acute{e}$s, \ Mart$\acute{i}$nez, \& Bullo}{2006}{rend} Cort$\acute{e}$s  J., \ Mart$\acute{i}$nez S., \&  Bullo F. \harvardyearleft2006\harvardyearright.
Robust rendezvous for mobile autonomous agents via proximity graphs
in arbitrary dimensions. {\em IEEE Trans. Autom. Control}, 51(8),
1289-1298.

\harvarditem{Smith, Broucke, \& Francis}{2007}{Franices}Smith S. L., \ Broucke M. E., \&  Francis B. \harvardyearleft2007\harvardyearright. Local control strategies for groups
of mobile autonomous agents. {\em IEEE Trans. Autom. Control},
52(6), 1154-1159.

\harvarditem{Pease, Shostak, \&  Lamport}{1980}{agree}Pease M., Shostak R., \& Lamport L. \harvardyearleft 1980\harvardyearright. Reaching agreement in the
presence of faults. {\em J. ACM}, 27, 228-234.


\harvarditem{Lobel, \& Ozdagar}{2011}{oz} Lobel H., \& Ozdagar A. \harvardyearleft 2011\harvardyearright. Distributed subgradient methods
for convex optimization over random networks. {\em  IEEE Trans.
Autom. Control}, 56(6), 1291-1306.



\harvarditem{Cao, Yu, \& Anderson}{2011} {formation} Cao M., Yu C., \& Anderson B. D. O. \harvardyearleft2011\harvardyearright. Formation control
using range-only measurements. {\em Automatica}, 47(4), 776-781.


\harvarditem{Jadbabaie, Lin, \& Morse}{2003}{Jad} Jadbabaie A., Lin J., \& Morse A. S. \harvardyearleft 2003\harvardyearright. Coordination of groups of mobile autonomous agents using nearest
neighbor rules. {\em IEEE Trans. Autom. Control}, 48(9), 988-1001.

\harvarditem{Ren, \& Beard}{2005}{ren} Ren W., \& Beard R. W. \harvardyearleft 2005\harvardyearright. Consensus seeking in multiagent systems under
dynamically changing interaction topologies.  {\em IEEE Trans. Autom. Control}, 50(5), 655-661.

\harvarditem{Olfati-Saber, \& Murry}{2004}{murry}
Olfati-Saber R., \&  Murray R.  \harvardyearleft 2004\harvardyearright. Consensus problems in networks of
agents with switching topology and time-delays, {\em IEEE Trans. Autom. Control}, 49 (9), 1520-1533.
\harvarditem{Moshtagh, Michael, Jadbabaie, \& Daniilidis}{2009}{vision1} Moshtagh N., Michael N., Jadbabaie A., and Daniilidis K. \harvardyearleft 2009\harvardyearright.
Vision-based, distributed control laws for motion coordination of
nonholonomic robots. {\em IEEE Trans. Robotics}, 25(4), 851-860.

\harvarditem{Montijano, Thunberg, Hu, \& Sag\"{u}\`{e}s}{2013}{vision2} Montijano E., Thunberg J., Hu X. M.,
\& Sag\"{u}\`{e}s C. \harvardyearleft2013\harvardyearright. Epipolar visual servoing for multirobot
distributed consensus. {\em IEEE Trans. Robotics}, 29(5),
1212-1225.

\harvarditem{Moreau}{2005}{moreau} Moreau L. \harvardyearleft2005\harvardyearright. Stability of multiagent systems with time-dependent communication
links. {\em IEEE Trans. Autom. Control}, 50(2), 169-181.

\harvarditem{Yu, Ren, Zheng, Chen, \& L$\ddot{u}}{2013}{yuwenwu} Yu W., Ren W., Zheng W., Chen G., \& L$\ddot{u}$ J. \harvardyearleft2013\harvardyearright.
Distributed control gains design for consensus in multi-agent
systems with second-order nonlinear dynamics. {\em Automatica},
49(7), 2107-2115.

\harvarditem{Xiao, \& Wang}{2008}{zhangya}Xiao F., \& Wang L. \harvardyearleft2008\harvardyearright. Consensus protocols for discrete-time
multi-agent systems with time-varying delays.  {\em Automatica},
44(10), 2577-2582.

\harvarditem{Wang, \& Liu}{2009}{wanglin}Wang L., \& Liu Z. X. \harvardyearleft2009\harvardyearright. Robust consensus of multi-agent systems with noise. {\em
Science in China: Information Science},  52(5), 824-834.

\harvarditem{Li, \& Zhang}{2009}{litao} Li T., \& Zhang J. F. \harvardyearleft2009\harvardyearright. Mean square average consensus under
measurement noises and fixed topologies: necessary and sufficient
conditions.  {\em Automatica}, 45(8), 1929-1936.
\harvarditem{Shi, \& Johansson}{2013}{shiguodong}Shi G. D., \& Johansson K. H. \harvardyearleft2013\harvardyearright. Robust consensus for continuous-time multi-agent dynamics.
{\em SIAM Journal on Control and Optimization}, 51(5),  3673-3691.
%

\harvarditem{Gupta, \& Kumar}{1999}{kumar}Gupta P., \& Kumar P. R. \harvardyearleft1999\harvardyearright.  Critical power for asymptotic connectivity in wireless networks. in
{\em Stochastic Analysis, Control, Optimization and Applications},
Birkhauser Boston, Boston, MA, 547-566.


\harvarditem{Ji, \& Egerstedt}{2007}{leader1}Ji M., \& Egerstedt M. \harvardyearleft2007\harvardyearright. Distributed coordination control of multiagent
systems while preserving connectedness.  {\em IEEE Trans.
Robotics}, 23(4), 693-703.

\harvarditem{Dimarogonas, \& Kyriakopoulos}{2007}{diamos}Dimarogonas D. V. ,\& Kyriakopoulos K. J. \harvardyearleft2007\harvardyearright. On the rendezvous problem for multiple
nonholonomic agents. {\em IEEE Trans. Autom. Control}, 52(5),
916-922.

\harvarditem{Dimarogonas, Tsiotras, \& Kyriakopoulos}{2009}{diamos1} Dimarogonasa D. V., Tsiotras P.,\& Kyriakopoulos K. J. \harvardyearleft2009\harvardyearright
Leader-follower cooperative attitude control of multiple rigid
bodies. {\em Systems and Control Letters}, 58, 429-435.
\harvarditem{Ajorlou, \& Aghdam}{2013}{connecprev}Ajorlou A.,\& Aghdam A. G. \harvardyearleft2013\harvardyearright Connectivity preservation in nonholonomic multi-agent
aystems: a bounded distributed control strategy,  {\em IEEE Trans.
Autom. Control}, 58(9), 2366-2371.

\harvarditem{Liu, Li, Xie, Fu, \& Zhang}{2013}{sampling1} Liu S., Li T., Xie L., Fu M.,\& Zhang J. F. \harvardyearleft2013\harvardyearright
Continuous-time and sampled-data based average consensus with
logarithmic quantizers.   {\em Automatica}, 49(11), 3329-3336.
\harvarditem{Xiao, \& chen}{2012}{sampling2}Xiao F., \& Chen T. \harvardyearleft2012\harvardyearright Sampled-data consensus for
multiple double integrators with arbitrary sampling. {\em IEEE
Trans. Autom. Control}, 57(12), 3230-3235.

\harvarditem{Tang, \& Guo}{2007}{tang2}Tang G. G., \& Guo L. \harvardyearleft2007\harvardyearright  Convergence of a
class of multi-agent systems in probabilistic framework.
 {\em Journal of Systems Science and Complexity}, 20(2), 173-197.
\harvarditem{Liu, \& Guo}{2009}{Auto}
Liu Z. X., \& Guo  L. \harvardyearleft2009\harvardyearright Synchronization of multi-agent systems
without connectivity assumption.  {\em Automatica}, 45(12),
2744-2753.

\harvarditem{Liu, Wang, \& Hu}{2014}{ifac} Liu  Z. X., Wang J. H., \& Hu  X. M. \harvardyearleft2014\harvardyearright Synchronization of Unicycle Robots with
Proximity Communication Networks. {\em Proc. of the 19th IFAC World
Congress}, 9197-9202, Cape Town, South Africa.

\harvarditem{Tove, Dimarogonas, Egerstedt, \& Hu}{2010}{toven}Tove G., Dimarogonas D. V., Egerstedt M., \&
Hu X. M.  \harvardyearleft2010\harvardyearright Sufficient conditions for connectivity maintenance and
rendezvous in leader-follower networks.   {\em Automatica}, 46(1),
133-139.
\harvarditem{Cao, Ren, \& Li}{2009}{renwei2}Cao Y., Ren W., \& Li Y. \harvardyearleft2009\harvardyearright Distributed
discrete-time coordinated tracking with a time-varying reference
state and limited communication. {\em Automatica}, 45(5),
1299-1305.
\harvarditem{Couzin, Krause,  Franks, \& Levy}{2005}{nature}Couzin I. D., Krause J., Franks N. R., \& Levy S.  \harvardyearleft2005\harvardyearright Effective leadership and
decision-making in animal groups on the move. {\em Nature}, 433,
513-516.
\harvarditem{Cao, Ren, \& Egerstedt}{2012}{leaders1}Cao Y. C., Ren W., \& Egerstedt  M. \harvardyearleft2012\harvardyearright Distributed containment control with
multiple stationary or dynamic leaders in fixed and switching
directed networks.  {\em Automatica}, 48(8), 1586-1597.

\harvarditem{Dimarogonas, Tsiotras, \& Kyriakopoulos}{2009}{leaders2} Dimarogonas D. V., Tsiotras P., \& Kyriakopoulos K. J. \harvardyearleft2009\harvardyearright
Leader-follower cooperative attitude control of multiple rigid
bodies.  {\em Systems \& Control Letters}, 58(6), 429-435.







\harvarditem{Penrose}{2003}{geometric}Penrose M. \harvardyearleft2003\harvardyearright Random geometric graphs.  {\em Oxford University Press}.


\harvarditem{Liu, Han, \& Hu}{2011}{auto2011}Liu Z. X., Han J., \& Hu X. M. \harvardyearleft2011\harvardyearright
The proportion of leaders needed for the expected consensus.
 {\em Automatica}, 47(12), 2697-2703.

\end{thebibliography}

\end{document}